\documentclass{aa}  

\usepackage{natbib}

\graphicspath{{./}{figures/}} 
\usepackage{txfonts}
\usepackage{gensymb}
\usepackage{placeins}
\usepackage{color}

\begin{document}

   \title{A 3D model for the stellar populations in the nuclei of NGC 1433, NGC 1566, and NGC 1808\thanks{The spectral library used in this paper is only available in electronic form
at the CDS via anonymous ftp to cdsarc.cds.unistra.fr (130.79.128.5)
or via https://cdsarc.cds.unistra.fr/cgi-bin/qcat?J/A+A/}}

   \subtitle{NIR photometry, CO absorption lines, and line-of-sight velocity and its dispersion}

   \author{
          P. Vermot
          \inst{1} 
          \and
          J. Palou\v s
          \inst{1}
          \and 
          B. Barna
          \inst{3}
          \and
          S. Ehlerov\' a
          \inst{1}
          \and 
          M. R. Morris
          \inst{2}
          \and
          R. W\" unsch
          \inst{1}
          }
     \institute{Astronomical Institute of the Czech Academy of Sciences,
              Bo\v{c}n\'{\i} II 1401/1, 140 00 Prague\\
              \email{pierre.vermot@asu.cas.cz}
              \and
           Department of Physics and Astronomy, University of California, Los Angeles, CA 90095-1547, USA
              \and
           Physics Institute, University of Szeged, D\'{o}m t\'{e}r 9, Szeged, 6723, Hungary
             }

   \date{Received September 15, 1996; accepted March 16, 1997}

 
  \abstract
   {}
   {We aim to characterize the properties of the stellar populations in the central few hundred parsecs of nearby galactic nuclei; specifically their age, mass, and 3D geometry.}
   {We use spatially resolved spectroscopic observations of NGC 1433, NGC 1566, and NGC 1808 obtained with SINFONI to constrain a 3D model composed of a spherically symmetric nuclear star cluster (NSC) and an extended thick stellar disk. We computed UV to mid-infrared single stellar population (UMISSP) spectra to determine the age of the stellar populations and construct synthetic observations for our model. To overcome degeneracies between key parameters, we simultaneously fit the spatially resolved line-of-sight velocity, line-of-sight-velocity-dispersion, low-spectral-resolution NIR continuum, and high-spectral-resolution CO absorption features for each pixel.}
   {For the three objects, we derive the age and mass of the young and old stellar populations in the NSC and surrounding disk, as well as their 3D geometry: radius for the NSC; thickness, inclination, and position angle for the disk. These results are consistent with published independent measurements when available. }
   {The proposed method allows us to derive a consistent 3D model of the stellar populations in nearby galactic centers solely based on a near-infrared IFU observation.}

   \keywords{ Galaxies: nuclei --
                 Galaxies: stellar content --
                 Galaxies: kinematics and dynamics
               }

   \maketitle
%

\section{Introduction}

In a series of papers \citep{Palous2020, Barna2022, Ehlerova2022}, we are using hydrodynamics simulations to investigate the potential of supernova
remnants to feed supermassive black holes (SMBH)  in the vicinity of the center of the Milky Way. We have discovered that nearby supernovae can have a positive impact on the growth of Sgr A* by pushing material from the interstellar medium (ISM) deeper into the central potential well.

We wish to expand this investigation to other galactic nuclei, whose properties are expected to vary (gravitational potential, ISM distribution, supernova rate), potentially leading to different results. For this purpose, we need a proper description of the stellar populations and ISM in these regions.

In this paper, we aim to determine the gravitational potential for three nearby galactic nuclei: NGC 1808, NGC 1433, and NGC 1566. These targets were chosen because they have been observed with both SINFONI and ALMA. The former is used in this paper to determine the gravitational potential, and the latter will be used in an upcoming publication to determine the ISM distribution. Recent publications on the same SINFONI data are associated with each of these objects, but none of them have provided a sufficiently detailed description of the gravitational potential:
 The analysis of NGC 1808 is presented in \citet{Busch2017}, where the authors fit a 2D circular Plummer model with an inclination given by the Sérsic profile of the galaxy to reproduce the stellar line-of-sight velocity (LOSV) map. However, the numerical values of the best parameters are not provided, as the fit is only used to subtract the circular motion from the map in order to detect noncircular motions of the gas and stellar content. A qualitative analysis of the LOSV dispersion ($\sigma_v$) is presented.
 NGC 1433 is described in \citet{Smajic2014}. The stellar LOSV is not fitted, but the authors present an analysis of the gaseous content, where a combination of disk-like rotation and a one-sided outflow is used to explain the measured LOSV.  
 The observation of NGC 1566 is presented in \citet{Smajic2015}, where the authors also fit a 2D circular Plummer model with a fixed inclination to reproduce the stellar LOSV map and qualitatively discuss the $\sigma_v$ map. The numerical values of the best parameters for the Plummer model are not provided, but it is indicated that no deviation from circular motion is detected. Analyzing the same data, we aim to improve the characterization of the stellar populations in these objects.

Two main issues can explain the limited description of the stellar mass distribution in previous publications: Firstly, there is degeneracy between mass, inclination, and age: when using photometry to determine the mass distribution, a strong uncertainty is associated with the unknown age of the stellar population, which can lead to differences in the mass-to-luminosity ratio  (M/L) of orders of magnitude; when using the LOSV distribution, a perfect degeneracy appears between the mass and the inclination of the disk. The second issue is the difficulty in determining the 3D mass distribution due to projection effects: it is particularly challenging to determine the geometrical thickness of a stellar disk with photometry and LOSV if not viewed edge-on.

In order to overcome these difficulties, we use a high-resolution single stellar population (SSP) library to determine the age of the stellar populations and perform a simultaneous fit of four observables determined for each pixel: a low-resolution near-infrared (NIR) spectral energy distribution (SED), a high-resolution absorption spectrum of the CO bandheads, LOSV, and $\sigma_v$. This allows us to break degeneracies, bring 3D information, and obtain reliable estimates of the main parameters of the model.

The targets and observations are presented in Section \ref{sec:obs}, the construction of the SSP spectra in Section \ref{sec:ssp}, the mass distribution model in Section \ref{sec:model}, the results of the fitting procedure in Section \ref{sec:results}, and a discussion and conclusion in Sections \ref{sec:disc} and \ref{sec:con}. Additional figures supporting the analysis are presented in the appendices of the paper.

\section{Observations}
\label{sec:obs}
\subsection{Targets and observations}

Three objects are analyzed in this paper: NGC 1808, NGC 1433, and NGC 1566. The method was developed on the first of these, for which it produces the best results, and was tentatively applied to the two other objects with good results.

The modeling presented in this paper is done solely on archival SINFONI data obtained in the high spectral resolution and medium spatial resolution modes. For each observation, the official ESO pipeline was used to perform the data reduction, and the flux calibration was performed on the entire field of view with the 2MASS extended source survey as a reference.

\textbf{NGC 1808} was observed as part of the 075.B-0648(A) ESO program on 24 March 2024, 2 April 2005 and 12 December 2005; each time in the three NIR J ($R\sim2000$), H ($R\sim3000$), and K ($R\sim4000$) spectral bands. A $6" \times 6"$ field of view is extracted with a $125 \times 250\ mas $ pixel scale and an effective resolution of 0.7". We assume a distance of $12.8\ \mathrm{Mpc}$, corresponding to $z \sim 0.003$ and $62\ \mathrm{pc/arcsec}$. A first analysis of this observation and a more detailed description of the host galaxy are presented in \citet{Busch2017}.

\textbf{NGC 1433} was observed as part of the 90.B-0657(A) ESO program on 21 October 2012 in the H and K bands. A $4" \times 4"$ field of view is extracted, with a $125 \times 250\ \mathrm{mas} $ pixel scale and an effective resolution of 0.6". We assume a distance of $15.5\ \mathrm{Mpc}$, corresponding to $75\ \mathrm{pc/arcsec}$. A first analysis of this observation and a more detailed description of the host galaxy are presented in \citet{Smajic2014}.

\textbf{NGC 1566} was observed as part of the 90.B-0657(A) ESO program on 21 October 2012 in the H and K bands. A $4.8" \times 4.8"$ field of view is extracted with a $125 \times 250\ \mathrm{mas} $ pixel scale and an effective resolution of 0.6".  We assume a distance of $21.56\ \mathrm{Mpc}$, corresponding to $105\ \mathrm{pc/arcsec}$. A first analysis of this observation and a more detailed description of the host galaxy are presented in \citet{Smajic2015}.

\subsection{The observables}
\label{subsec:observables}
For each of the objects mentioned above, we extract the following information:
\begin{enumerate}
    \item A low-spectral-resolution data cube, where the angular resolution, field of view, and spectral coverage of the observations are preserved, but the spectral resolution is lowered by a factor 50 by rebinning the original data cube in the spectral dimension. As the continuum emission is dominated by young stellar sources, this provides a spatially resolved SED for each pixel that will constrain the age, mass, and spatial distribution of the young stellar populations around the nuclei.
    \item A high-spectral-resolution data cube, where the angular resolution, field of view, and spectral resolution of the observations are preserved, but the spectral domain is restrained to $ 2.25-2.40 \mathrm{\mu m} $ (after correction of the redshift) and the flux normalized by the continuum. This provides the absorption profile of the CO bandheads for each pixel, which can be used to constrain the age and relative importance of the old and young stellar populations.
    \item An LOSV map, obtained by fitting the above-mentioned absorption features with the high-spectral-resolution SSP spectra described below. For each pixel, we estimate the velocity shift as the median of 30 Doppler shifts measured with SSPs of randomly chosen ages. We use the LOSV map to constrain the total mass distribution of our model.
    \item A $\sigma_v$ map obtained with the same procedure as in point (3). We use this measurement to constrain the thickness of the stellar disk and solve the degeneracy between mass and inclination. 
\end{enumerate}

The simultaneous use of these four sources of information to constrain our models will bring much stronger constraints than if fitted individually.
In addition to these observables, the flux from the [Fe II] emission at 1.64 $\mathrm{\mu}$m is measured in each pixel as the integrated emission line after a linear continuum subtraction.

\section{Stellar templates}
\label{sec:ssp}
To determine the age of the stellar populations and the LOSV and $\sigma_v$ maps, we need a high-spectral-resolution library of unresolved stellar populations at various ages. Several spectral libraries computed
with evolutionary SSP models are publicly available. Such models are built with an initial mass function, stellar evolution tracks, and spectral templates for individual stars, offering a high degree of freedom. 

Despite the relatively large number of possible libraries to choose from, we found no public stellar library matching our requirements, which are:
\begin{enumerate}
    \item Solar metallicity. We lack sufficient information to simultaneously fit the ages and metallicities of the stellar populations, and therefore we assume a solar metallicity for the three objects as our best estimate for an elliptical star-forming galactic center.
    \item Coverage of the NIR J, H, and K bands to match the spectral range of SINFONI.
    \item Spectral resolution of $R \gtrsim 3000$ ($\Delta v \lesssim 100\ \mathrm{km.s^{-1}}$) in the NIR to fit the shape of the absorption features and measure LOSV and $\sigma_v$.
    \item Spectra for very young post-starburst populations down to 10 Myr to investigate regions actively producing supernov\ae.
\end{enumerate}

Several public SSP libraries fulfill some of these criteria; MaStar \citep{Maraston2020}, A-LIST \citep{Ashok2021}, MILES \citep{Vazdekis2016}, XSL \citep{Verro2022}, and GALEV \citep{Kotulla2009} being the most notable ones. MaStar, which is based on the MaNGA Stellar Library, is constructed with the empirical spectra of approximately $9000$ SDSS stars with a very wide range of parameters. However, it only covers the shortest NIR wavelengths (up to 1.03 $\mathrm{\mu m}$), does not provide sufficient spectral resolution to fit the $\sigma_v$, and does not cover ages younger than 200 Myr.  The APOGEE Library of Infrared SSP Templates (A-LIST) provides high-spectral-resolution templates, but only in a narrow band (1.51–1.70 $\mathrm{\mu m}$) and for stellar populations older than 2 Gyr. 
    
    The MILES extended SSP model is based on a composite library of empirical spectra covering a wide spectral range (NGSL for the UV, MILES for the optical, IRTF for the infrared), with moderate to high spectral resolution. Unfortunately, its spectral resolution in the NIR is R$\sim2000-2500$, corresponding to $\Delta \lambda \sim 1 \mathrm{nm}$, or $\Delta v \sim 150\ \mathrm{km.s^{-1}}$, which is insufficient to fit stellar $\sigma_v$. Although it provides spectra for stellar populations as young as 30 Myr, which is close to our goal. 
    
    The X-shooter Spectral Library (XSL) from \citep{Verro2022}, which is based on empirical spectra from the eponym instrument, provides spectra at high resolution (R $\sim$ 10 000) over a wide spectral range (350–2480 nm) and provides spectra for populations as young as 50 Myr. Lastly,  the GALaxy EVolutionary (GALEV) synthesis models are fast, publicly available models for the computation of SSP spectra. These can be used to predict spectra for very young ages, down to 4 Myr, and use the theoretical Basel Stellar Library to cover a very large spectral domain from the extreme UV to the far-infrared (FIR). However, their spectral resolution is rather low ($\Delta \lambda \sim 5-10\ \mathrm{nm}$ in the NIR) and therefore they are not suitable for the analysis of absorption features.

Unable to find a spectral library matching our requirements, we constructed our own using the Stellar Population Interface for Stellar Evolution and Atmospheres \citep[SPISEA, ][]{Hosek2020}. Our library matches and surpasses the above criteria, notably thanks to the use of synthetic spectra obtained with the stellar atmosphere code PHOENIX \citep{Allard2003, Allard2007, Allard2012}. As the construction of our library required a significant amount of resources and can now be rapidly used for the analysis of other observations, we have chosen to make it publicly available.


\subsection{Initial mass function}

We use a standard Kroupa IMF \citep{Kroupa2001}, which is a double-break power law, defined as
\begin{equation}
    \xi (m) \propto m^{{-\alpha }}
    \begin{cases}
      \alpha = 1.3\text{ for } 0.1 < m < 0.5\\
      \alpha = 2.3\text{ for } 0.5 < m < 300\\
    \end{cases}\,,
\end{equation}
to generate an initial population of stars. No stellar multiplicity is introduced.

Due to the stochastic construction of the initial population, there is variability in the final spectra. For this reason, we compute ten $10^6\ \mathrm{M_{\odot}}$ models for each age, providing good precision on the average spectrum as well as an estimate of the variability at a given age. The standard deviation of each spectrum is provided in the library.

\subsection{Stellar evolution}
For each age, an isochrone is computed to be used to attribute physical properties ($T_{eff}$, $log\ g$, and $L$) to each star of the initial population.
We chose to use the MESA isochrones and stellar tracks \citep[MIST; ][]{Dotter2016, Choi2016, Paxton2011, Paxton2013, Paxton2015}, whose evolutionary tracks are computed with the MESA 1D stellar evolution code and cover a wide range of parameters: stellar mass (including high-mass stars, [0.1; 300]$\ \mathrm{M_\odot}$), age (including very young stars, [$10^6$; $10^{10}]\ $yr), and metallicity ([Fe/H], [-4$; $ 0.5]). We used the version defined as v1.2 in SPISEA, which was downloaded in August 2018 (solar metallicity) and in April 2019 (other metallicities).

Each star is then attributed a synthetic spectrum from the PHOENIX Models for Synphot\footnote{Described and available for download at \url{https://www.stsci.edu/hst/instrumentation/reference-data-for-calibration-and-tools/astronomical-catalogs/phoenix-models-available-in-synphot}} \citep{Allard2003, Allard2007, Allard2012}, which cover the 50-50000 nm spectral range. For [Fe/H]=0, the library is at high spectral resolution ($R \sim 14000$) and was constructed with no missing template except for the Wolf-Rayet stars.

This model was used to compute a library that we call UMISSP ({UV to mid-infrared single stellar population}), which consists of synthetic spectra of $10^6\ \mathrm{M_{\odot}}$ solar-metallicity stellar populations with ages ranging from $10^6$ to $10^{10}$ yr with $10^{0.05}$ steps\footnote{The library is publicly available at \url{http://galaxy.asu.cas.cz/page/umisspr}. Spectra at lower resolution ($R \sim 280$) are also computed for other metallicities ([Fe/H] = 0.5, -0.5, -1, -2, -3, -4), although not used in this paper. The mean spectra and standard deviation spectra for all ages and metallicities are provided. They are flux calibrated and expressed in $\mathrm{W.\mu m^{-1}}$, corresponding to the total cluster emission in a 4$\pi$ sr solid angle. In addition, for each stellar population, we provide: (a) the initial mass histogram, (b) the MIST isochrone, (c) the parameters of the isochrone and the ones used for the spectrum attribution of each star.}. In Appendix \ref{app:ssp}, we perform several basic tests to verify that our spectra are in general agreement with other SSP models.

\section{Model}
\label{sec:model}
In order to model the observations, we choose a two-component model composed of a spherical nuclear cluster and an extended thick disk. For each set of parameters, we construct 3D maps of the various observables, which we stack along the line of sight. These are then used to compute synthetic observations, which are directly compared to the above-mentioned set of observables (Subsection \ref{subsec:observables}). The construction of these synthetic observations is described in more detail in the following paragraphs.
\subsection{3D maps of mass, velocity, and velocity dispersion}
We first construct 3D maps of the mass distribution, circular velocity, and velocity dispersion for each component using two self-consistent potential density function pairs: the Plummer spherical cluster \citep{Plummer1911}  and the Miyamoto-Nagai thick disk \citep{Miyamoto1975}.

The mass distribution is computed as

\begin{equation}
\rho_{NSC}(R, z) = \frac{3M_{NSC}}{4\pi b_{NSC}^3} \left(1 + \frac{R^2+z^2}{b_{NSC}^2}\right)^{-\frac{5}{2}}, 
\end{equation}
for the nuclear star cluster (NSC) Plummer model, and
\begin{equation}
\begin{aligned}
&\rho_{disk}(R, z) = \\
&\frac{b^2 M}{4\pi}\frac{a_{MN} R^2+[a_{disk}+3(z^2+b_{disk}^2)^{1/2}][a_{disk}+(z^2+b_{disk}^2)^{1/2}]^2}{(R^2+[a_{disk}+(z^2+b_{disk}^2)^{1/2}]^2)^{5/2}(z^2+b_{disk}^2)^{3/2}}
\end{aligned}
\end{equation}
for the extended Miyamoto-Nagai disk.

The corresponding potentials are

\begin{equation}\Phi_{NSC}(R, z) = -\frac{G M_{NSC}}{\sqrt{R^2 + z^2 + b_{NSC}^2}}, 
\end{equation}

\begin{equation}\Phi_{disk}(R, z) = -\frac{G M_{disk}}{\sqrt{R^2+[a_{disk}+(z^2+b_{disk}^2)^{1/2}]^2}}, \end{equation}
 where (R, $\theta$, z) are the cylindrical coordinates in the disk.

The resulting total gravitational potential is

\begin{equation}\Phi(R, z) = \Phi_{NSC}(R, z) + \Phi_{disk}(R, z). \end{equation}

We assume that the central cluster is spherical and has a rest-frame LOSV. For the extended disk, we compute the circular velocity as the derivative of the total gravitational potential (NSC + disk):

\begin{equation}
\overrightarrow{v}_{disk}(R, z) = \sqrt{r\frac{d\Phi}{dr}}\left(\frac{R}{r}\right) \overrightarrow{e}_{\theta}, 
\end{equation}
where $r = \sqrt{R^2+z^2}$ and $\overrightarrow{e}_{\theta}$ is the unit vector in the direction of rotation.

For the NSC Plummer cluster \citep{Dejonghe1987}, the local isotropic velocity dispersion is given by 

\begin{equation}
\sigma_{NSC}^2(r) = \frac{G M_{NSC}}{6\sqrt{r^2 + b_{NSC}^2}}, 
\end{equation}
 while for the extended MN disk, we assume that the local velocity dispersion is isotropic and equal to the product of the orbital velocity by the aspect ratio $(h/r)^{*}$ of the disk:

\begin{equation}
\sigma_{disk}(R, z) = (h/r)^{*}\sqrt{r\frac{d\Phi_{NSC+disk}}{dr}}\left(\frac{R}{r}\right), 
\end{equation}

where $(h/r)^{*}$ is the scale height of the disk, that is, the ratio between the vertical and radial spatial scales. We note that this ratio is not equal to $\frac{b_{disk}}{a_{disk}}$, but was measured by fitting an elongated Gaussian on $\rho_{disk}$ distributions for a wide range of $(a_{disk}, b_{disk})$ parameters, and is interpolated at the desired values (see Fig. \ref{app:mndisk}).

In the end, the LOSV of each component is given by

\begin{equation}
    \begin{cases}
      v_{disk, los}(R, z) = \overrightarrow{v}_{disk}(R, z) \cdot \overrightarrow{e}_{los} = v_{disk}(R, z) \times sin(i)sin(\theta) \\
      v_{NSC, los}(R, z) = 0, \\
    \end{cases}
\end{equation}
where $\overrightarrow{e}_{los}$ is the unit vector along the direction of the LOS, $i$ is the inclination of the disk (null for a face-on disk), and $\theta$ the azimuth angle in the disk (null in the direction of the observer).

\subsection{Spectral templates}

With the model described above, we compute 3D grids for each component, with a mass, LOSV, and LOSV dispersion attributed to each cell. We use this information to attribute a spectrum to each cell of the grid, as follows:
\begin{enumerate}
    \item Based on the high-resolution SSP spectra presented above, we attribute two spectra to each component: a low-resolution component with the full spectral domain, and a high-resolution component limited to the $2.25-2.40\ \mathrm{\mu m}$ wavelength interval. Both of them are scaled according to mass enclosed in each cell.
    \item We perform a Doppler shift of the high-resolution spectrum based on the projected velocity along the LOS.
    \item We convolve the high-resolution spectrum with a Gaussian whose width is given by the $\sigma_v$.
    \item We convolve the high-resolution spectrum with the SINFONI line spread function (measured on strong OH lines from the SKY observation, which are assumed to be constant within the Field of View FoV).
\end{enumerate}

\subsection{Synthetic observations}

At last, we compute the synthetic observation corresponding to our four observables:
\begin{enumerate}
    \item A low-resolution datacube obtained by summing the low-resolution spectra along the LOS.
    \item A high-resolution datacube obtained by summing the shifted and convolved high-resolution spectra along the LOS, and normalizing it to the measured continuum.
    \item A LOSV map obtained by calculating the average of the velocity projected along the LOS weighted by the luminosity of each cell at 2.3 $\mathrm{\mu m.}$
    \item A LOSV dispersion map obtained by computing the square root of the quadratic sum of (i) the average of the local velocity dispersion projected along the LOS, and (ii) the standard deviation of the velocity projected along the LOS, with both weighted by the luminosity in each cell at 2.3 $\mathrm{\mu m}$.
\end{enumerate}

All the synthetic observations are convolved with a Gaussian kernel matching the angular resolution of the observation.

\section{Results}
\label{sec:results}
\subsection{Fitting procedure}

The model consists of an NSC with two stellar populations ---one young and one old--- characterized by their masses $M_{Y,NSC}$ and $M_{O, NSC}$ and ages $age_{Y, NSC}$ and $age_{O, NSC}$, respectively; and similarly for the disk with $M_{Y,disk}$ and $M_{O, disk}$ and ages $age_{Y, disk}$ and $age_{O, disk}$. The two stellar populations of each component are assumed to be extinguished by the same amount of foreground material, which are defined by $A_{K, NSC}$ and $A_{K, disk}$. Additionally, the geometry of the cluster is given by only one parameter, its radius $b_{NSC}$; while the geometry of the disk is described by four parameters, namely its radius $a_{disk}$, its scale height $(h/r)^*$, its inclination $i,$ and its position angle $PA$. Overall, the model has 15 free parameters, and the computation time for each set of parameters is between 100 ms and 1000 ms. This prevents us from fully exploring the parameter space, and requires the use of a specific two-step  fitting procedure. We first determine the eight parameters corresponding to the ages of the stellar populations, namely $age_{Y, NSC}$, $age_{O, NSC}$, $M_{Y, NSC}/M_{O, NSC}$, $A_{K, NSC}$, $age_{Y, disk}$, $age_{O, disk}$, $M_{Y, disk}/M_{O, disk}$, and $A_{K, disk}$, as follows:
\begin{enumerate}
    \item We extract integrated spectra of the observed NSC and extended disk with aperture spectrometry. The average disk spectrum is subtracted from the NSC one (masks are determined based on the continuum images, where the NSC is clearly distinguishable).
    \item For each component (NSC and disk), we measure one low-resolution spectrum covering the JHK bands, and a high-resolution spectrum covering the CO bandheads at 2.4 $\mathrm{\mu m}$. 
    \item For each component, we fit an extinction $A_K$ and the relative mass contribution ($M_{Y}/M_{O}$) between two stellar populations with different ages. We try every combination of young ($10^6\ \mathrm{yr} \leq age_Y \leq 10^{9}\ \mathrm{yr}$ in steps of $10^{0.025}$) and old ($10^{9}\ \mathrm{yr} \leq age_O \leq 10^{10}\ \mathrm{yr}$ in steps of $10^{0.1}$) stellar populations, measure the sum of the squared residuals both on the observed low-resolution continuum and on the normalized high-resolution absorption spectrum (weighted by the inverse of the average value of each of these), and keep the best combination for the second step of the fit (secondary solutions have also been tried). 
\end{enumerate}
Once we have fixed the ages of the two stellar populations, we fit the masses and geometrical parameters, $M_{NSC}$, $b_{NSC}$, $M_{disk}$, $a_{disk}$, $b_{disk}$, $i$ and $PA$:
\begin{enumerate}
    \item At first, we find an initial solution by manually adjusting the parameters and visually inspecting the synthetic observables, as follows:
    \begin{itemize}
        \item $M_{NSC, 0}$ and $r_{NSC, 0}$ are set to reproduce the continuum images and the $\sigma_v$ map.
        \item $M_{disk, 0}$ and $a_{disk, 0}$ are chosen to approximately match the continuum images. 
        \item $PA_0$ is easily obtained by the line of nodes of the LOSV map.
        \item Keeping the previous estimate for $M_{disk, 0}$ and $a_{disk, 0}$, we fit $i_0$ and $b_{disk, 0}$ to best reproduce the observed LOSV and $\sigma_v$. 
    \end{itemize}
    \item We use the previous solution ($M_{NSC,0}$, $r_{NSC,0}$, $M_{disk,0}$, $a_{disk,0}$, $b_{disk,0}$, $i_0$, $PA_0$) as an initial guess for the automated Trust Region Reflective least-square algorithm implemented in \textit{scipy.optimize}, which provides the optimal solution around the initial guess. For each object, the weights associated to the four observables (continuum, CO absorption features, LOSV, and $\sigma_v$) are determined iteratively to give them similar importance in the fitting procedure. 
    \item Finally, we estimate the uncertainties on the parameters by exploring each of them individually, except for $M_{disk}$ and $i,$ which can be strongly correlated and for which a 2D map of the summed squared residuals is computed to determine the uncertainties (see Figs. \ref{fig:residuals_1808}, \ref{fig:residuals_1433}, and \ref{fig:residuals_1566}).
\end{enumerate}

The best parameters for the three objects are presented in the following tables, and the maps associated to the best models are presented in the appendices.

\subsection{Supernova rate}

\begin{figure}
    \centering
    \includegraphics[width=0.99\linewidth]{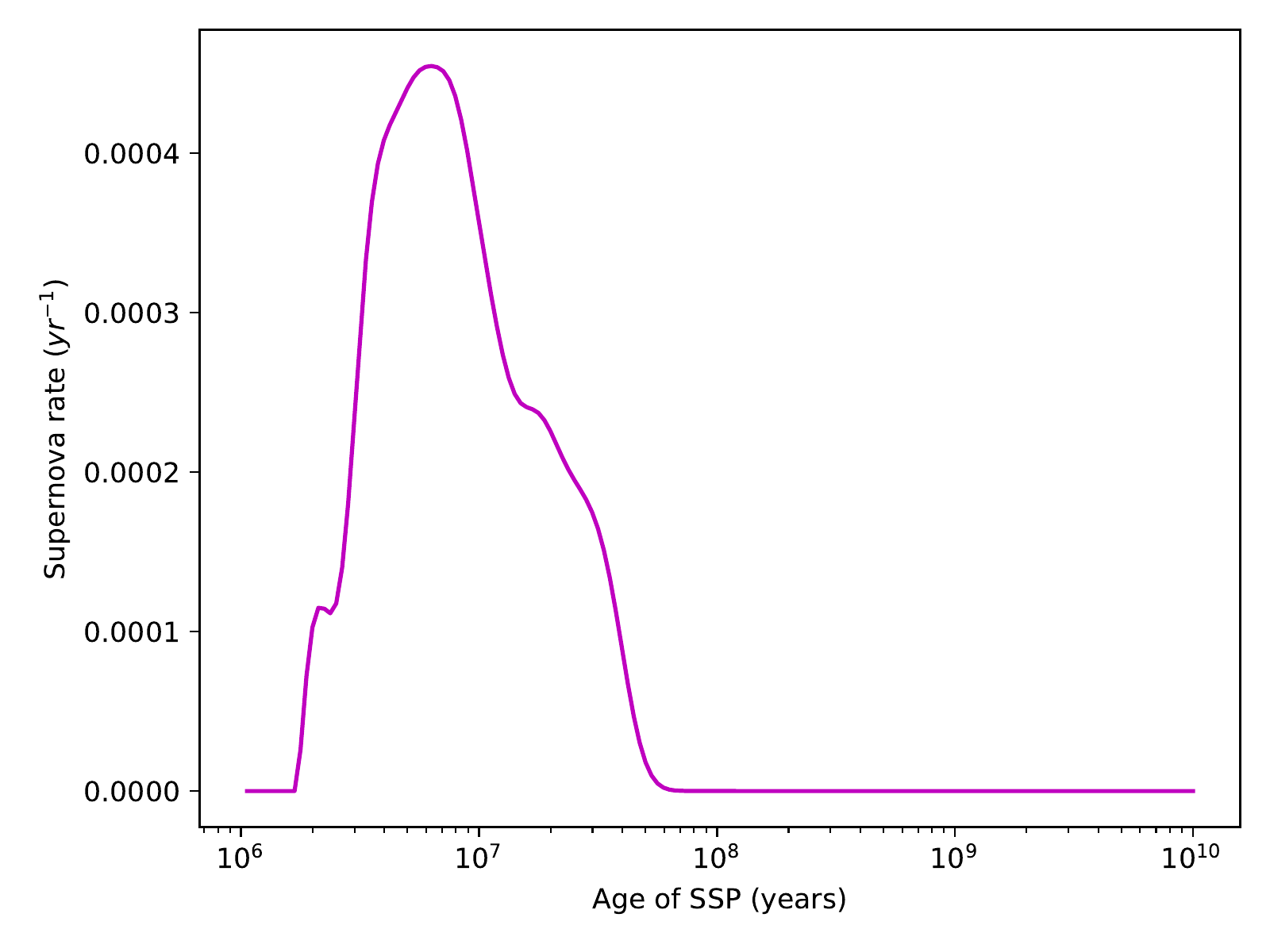}
    \caption{Evolution of the UMISSP supernova rate}
    \label{fig:snr_rate}
\end{figure}

With MIST, we can estimate the spatial distribution of the average supernova rate in each object. For a SSP at a given age, the specific supernova rate can be estimated by counting the number of stars with $M > 8 \mathrm{M_\odot}$ (smoothed over three temporal bins) and taking its derivative (see Fig. \ref{fig:snr_rate}). This simplified model assumes that all stars with $M > 8 \mathrm{M_\odot}$ will eventually explode as supernov\ae , and does not take into account type I.  We use this information and the mass distribution of each stellar population to construct a 3D map of the supernova rate.

Another way to estimate the average supernova rate is to measure the [Fe II] emission line flux, because the flux of these lines is observed to be strongly correlated in integrated galaxy spectra and on a pixel-to-pixel basis \citep{Rosenberg2012}:
\[\log \frac{v_{\mathrm{SNrate}}}{\mathrm{yr}^{-1} \mathrm{pc}^{-2}}=(1.01 \pm 0.2) * \log \frac{[\mathrm{FeII}]_{1.26}}{\mathrm{erg}.\mathrm{s}^{-1} \mathrm{pc}^{-2}}-41.17 \pm 0.9.\]

We also use this scaling relation to compute 2D supernova rate maps.
A similar relation is presented by \citet{Alonso2003}, which is based on the number of supernova remnant radio counts in M82 and NGC 253:

\[v_{\mathrm{SNrate}}= 0.8 \frac{L_{\mathrm{[Fe II]}}}{W}\times 10^{-34} \mathrm{yr}^{-1}.\]

\subsection{NGC 1808}

\begin{table}
\caption{NGC 1808: Results of the age-retrieval procedure}
\begin{center}
\begin{tabular}{c|c c}
\hline
\hline
log($age_{Y, NSC}$) & 7.2 & log(Myr) \\
\hline
log($age_{O, NSC}$) & 9.0 & log(Myr) \\
\hline
$M_{Y,NSC}/M_{O,NSC}$ & 0.10 &  \\
\hline
$A_{K, NSC}$ & 0.30 &  \\
\hline
log($age_{Y, disk}$) & 7.2 & log(Myr) \\
\hline
log($age_{O, disk}$) & 9.5 & log(Myr) \\
\hline
$M_{Y,disk}/M_{O,disk}$ & 0.04 &  \\
\hline
$A_{K, disk}$ & 0.20 &  \\
\hline
\end{tabular}
\end{center}
\end{table}

\begin{table}
\caption{NGC 1808: Mass and geometry parameters}
\begin{center}
\begin{tabular}{c|c c}
\hline
\hline
$M_{NSC}$ & 8.4 $\pm$ \ 0.2 & $log(\mathrm{M_{\odot}})$ \\
\hline
$M_{disk}$ & 8.5 $\pm$ \ 0.3 & $log(\mathrm{M_{\odot}})$ \\
\hline
$i$ & 46.3 $\pm$ \ 13.5 & $\degree$ \\
\hline
$PA$ & 233.6 $\pm$ \ 40.9 & $\degree$ \\
\hline
$b_{NSC}$ & 20.0 $\pm$ \ 5.0 & pc \\
\hline
$a_{disk}$ & 23.5 $\pm$ \ 5.9 & pc \\
\hline
$(h/r)_{disk}$ & 0.67 $\pm$ \ 0.07 &  \\
\hline
\end{tabular}
\end{center}
\end{table}

\begin{figure}[h]
    \centering
    \includegraphics[width=\linewidth]{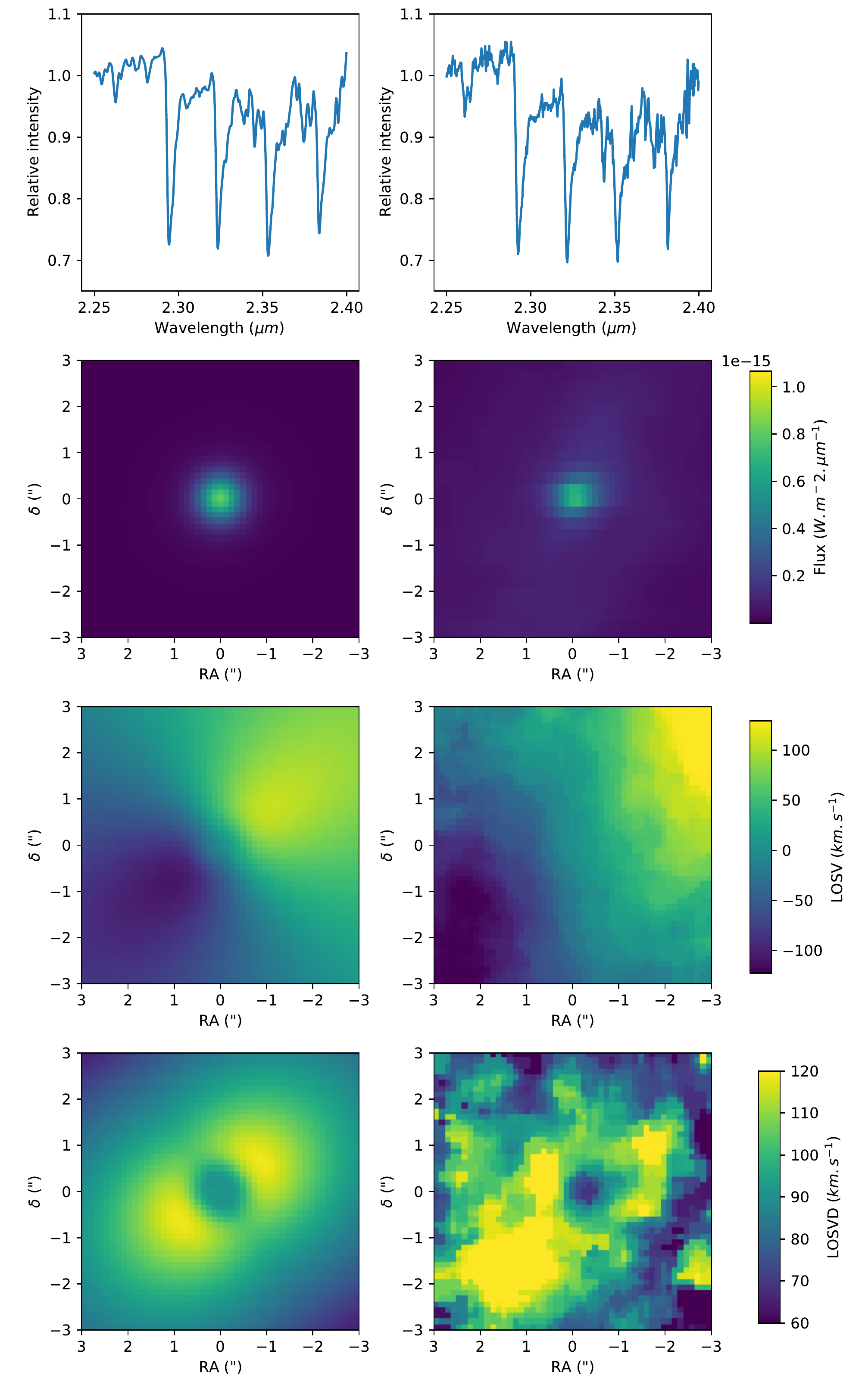}
    \caption{NGC 1808: Observation (right) and best model (left). From top to bottom: Averaged CO absorption lines over the entire FoV, H band image, LOSV, and LOSV dispersion.}
    \label{fig:check_3_11808}
\end{figure}

The nucleus of NGC 1808 appears to have undergone a recent starburst event, as evidenced by the estimated ages of young stellar populations in the disk and the NSC of $\sim 30\ \mathrm{Myr}$. Analysis of residuals in the continuum and CO absorption features (see Figs. \ref{fig:k2_age_ak_NSC_1808} and \ref{fig:k2_age_ak_disk_1808}) shows that the extinction is well constrained to a low value and several ages provide a good fit, with the deepest and widest minimum around the selected ages. We find similar ages for the populations in the disk and in the NSC, which supports the idea of a common star formation history. Two secondary solutions are observed at $10^{8.3}$ and $10^{8.5}$ yr, corresponding to the late apparition of red giants from $2-3.5\ \mathrm{M_{\odot}}$ stars, which briefly make the NIR spectrum similar to that of a young stellar population. 

A large NSC surrounded by a compact thick disk inclined by $i \sim 45 \degree$ and oriented along $PA \sim 230 \degree$ is the best model for NGC 1808. Although a larger disk radius and mass can improve the LOSV map fit, this leads to a poor fit for the $\sigma_v$ map and a slightly poorer fit for the continuum images. It is possible that two disk-like structures coexist, but for modeling the inner region, we focus on the compact disk solution.

The model offers a relatively good fit to all observations, with some noticeable differences. The colors and morphology of the central bright continuum source match the observations, but its size is hard to estimate due to limited angular resolution. The CO absorption features are well represented. The LOSV is slightly underestimated, and our model predictions suggest its maximum is closer to the nucleus than actually observed, indicating that it is missing mass at large radii. However, the model correctly predicts the position of the maximum of $\sigma_v$, as well as the central drop and its average value.

\subsection{NGC 1433}

\begin{table}
\caption{NGC 1433: Results of the age-retrieval procedure}
\begin{center}
\begin{tabular}{c|c c}
\hline
\hline
log($age_{Y, NSC}$) & 8.6 & log(Myr) \\
\hline
log($age_{O, NSC}$) & 9.8 & log(Myr) \\
\hline
$M_{Y,NSC}/M_{O,NSC}$ & 0.11 &  \\
\hline
$A_{K, NSC}$ & 0.60 &  \\
\hline
log($age_{Y, disk}$) & 8.2 & log(Myr) \\
\hline
log($age_{O, disk}$) & 9.8 & log(Myr) \\
\hline
$M_{Y,disk}/M_{O,disk}$ & 0.04 &  \\
\hline
$A_{K, disk}$ & 0.25 &  \\
\hline
\end{tabular}
\end{center}
\end{table}

\begin{table}
\caption{NGC 1433: Mass and geometry parameters}
\begin{center}
\begin{tabular}{c|c c}
\hline
\hline
$M_{NSC}$ & 9.3 $\pm$ \ 0.4 & $log(\mathrm{M_{\odot}})$ \\
\hline
$M_{disk}$ & 10.4 $\pm$ \ 0.25 & $log(\mathrm{M_{\odot}})$ \\
\hline
$i$ & 2.3 $\pm$ \ 2.2 & $\degree$ \\
\hline
$PA$ & 335.1 $\pm$ \ 54.5 & $\degree$ \\
\hline
$b_{NSC}$ & 35.8 $\pm$ \ 8.9 & pc \\
\hline
$a_{disk}$ & 200.0 $\pm$ \ 17.5 & pc \\
\hline
$(h/r)_{disk}$ & 0.23 $\pm$ \ 0.06 &  \\
\hline
\end{tabular}
\end{center}
\end{table}
\begin{figure}[h]
    \centering
    \includegraphics[width=\linewidth]{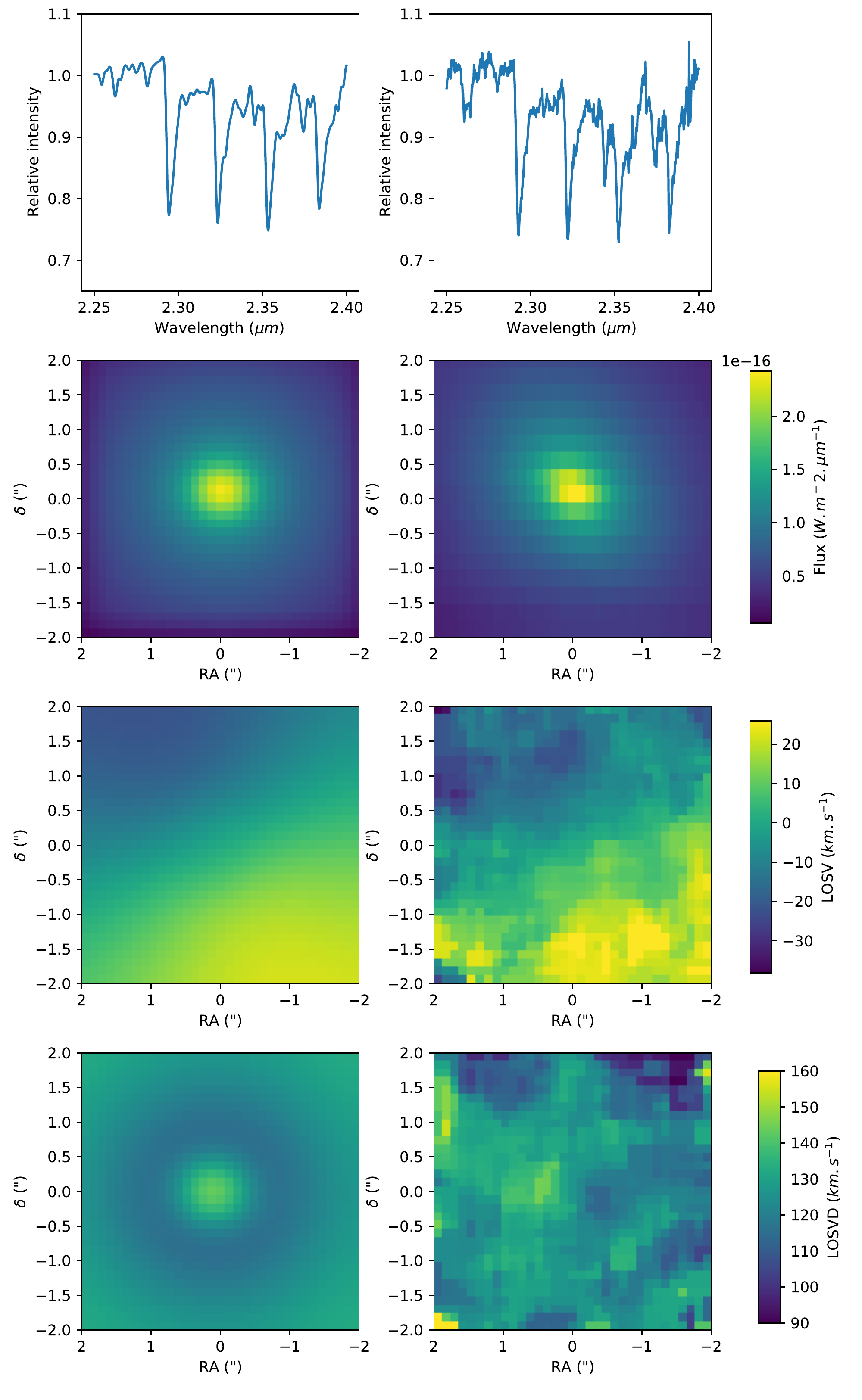}
    \caption{NGC 1433: Observation (right) and best model (left). From top to bottom: Averaged CO absorption lines over the entire FoV, H band image, LOSV, and LOSV dispersion.}
    \label{fig:check_3_1433}
\end{figure}

NGC 1433 shows no evidence of recent star formation, with the youngest populations estimated to be over 100 million years old in both the disk and NSC. The spectra of both components are similar, and so are the residual maps in Figures \ref{fig:k2_age_ak_NSC_1433} and \ref{fig:k2_age_ak_disk_1433}, despite the NSC being more extincted.

The best model features a massive NSC of $\gtrsim 10^9\ \mathrm{M_\odot}$ with a radius of $\sim 25$ pc surrounded by an extended, almost face-on disk of $\sim 300\ \mathrm{pc}$ in diameter with a height-to-radius ratio of $\sim 0.25$ and an inclination of $\lesssim 10\degree$.

The model offers a fair representation of all observables, particularly the CO absorption lines, continuum images, and LOSV map, which match the observations well. The $\sigma_v$ map geometry is less convincing, but is still reasonable: the observed $\sigma_v$ map is mostly featureless, and the model accurately reproduces its average value, which is relatively high. The model would predict a slight drop around the NSC, but this is not observed.




\subsection{NGC 1566}
\begin{table}
\caption{NGC 1566: Results of the age-retrieval procedure}
\begin{center}
\begin{tabular}{c|c c}
\hline
\hline
log($age_{Y, NSC}$) & 6.6 & log(Myr) \\
\hline
log($age_{O, NSC}$) & 9.8 & log(Myr) \\
\hline
$M_{Y,NSC}/M_{O,NSC}$ & 0.08 &  \\
\hline
$A_{K, NSC}$ & 0.90 &  \\
\hline
log($age_{Y, disk}$) & 6.1 & log(Myr) \\
\hline
log($age_{O, disk}$) & 9.2 & log(Myr) \\
\hline
$M_{Y,disk}/M_{O,disk}$ & 0.10 &  \\
\hline
$A_{K, disk}$ & 0.40 &  \\
\hline
\end{tabular}
\end{center}
\end{table}

\begin{table}
\caption{NGC 1566: Mass and geometry parameters}
\begin{center}
\begin{tabular}{c|c c}
\hline
\hline
$M_{NSC}$ & 8.5 $\pm$ \ 0.4 & $\mathrm{log(\mathrm{M_{\odot}})}$ \\
\hline
$M_{disk}$ & 10.1 $\pm$ \ 0.3 & $\mathrm{log(\mathrm{M_{\odot}})}$ \\
\hline
$i$ & 5.4 $\pm$ \ 5.0 & $\degree$ \\
\hline
$PA$ & 322.1 $\pm$ \ 40.3 & $\degree$ \\
\hline
$b_{NSC}$ & 7.9 $\pm$ \ 2.0 & pc \\
\hline
$a_{disk}$ & 110.0 $\pm$ \ 28.9 & pc \\
\hline
$(h/r)_{disk}$ & 0.22 $\pm$ \ 0.05 &  \\
\hline
\end{tabular}
\end{center}
\end{table}

\begin{figure}[h]
    \centering
    \includegraphics[width=\linewidth]{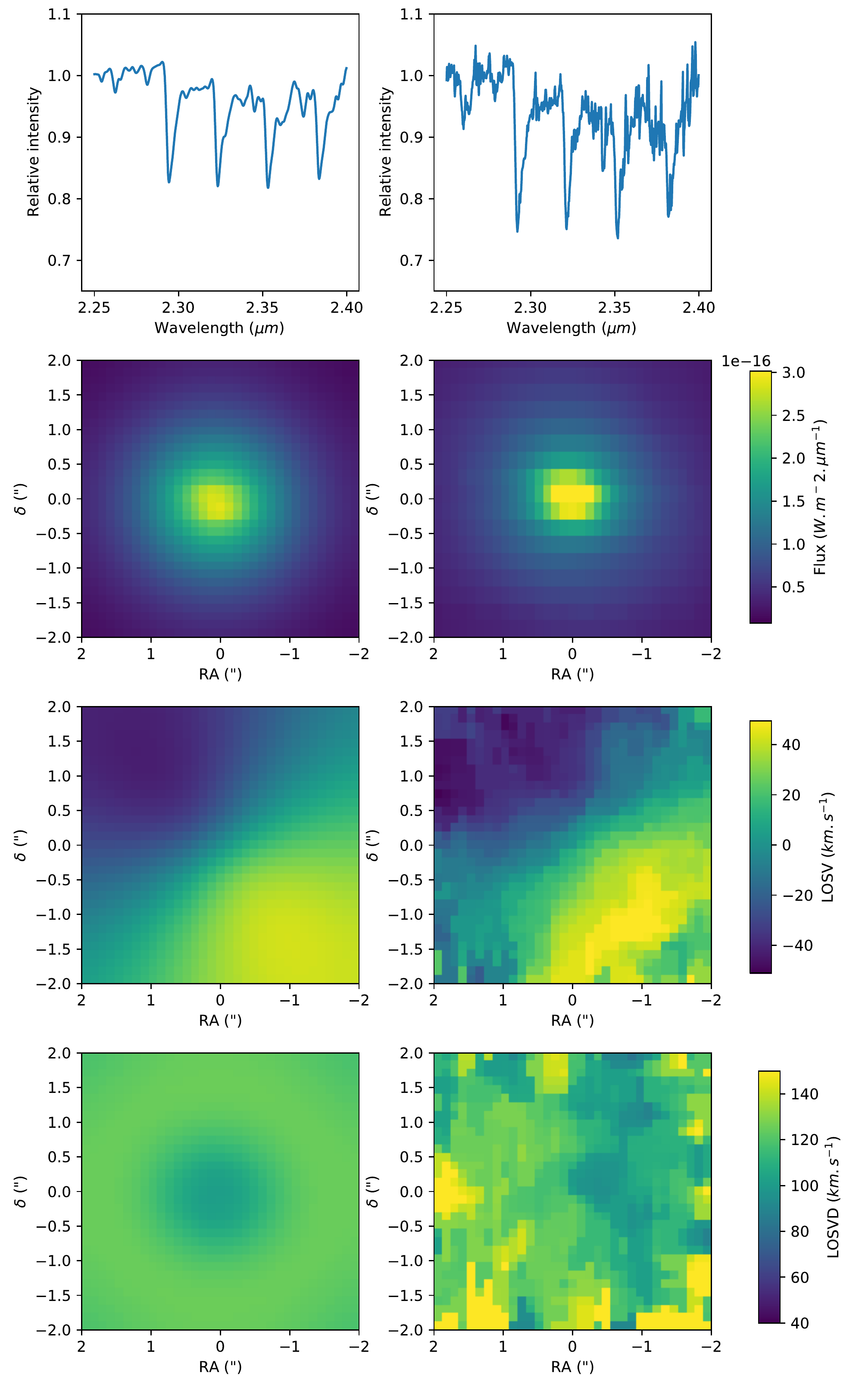}
    \caption{NGC 1566: Observation (right) and best model (left). From top to bottom: Averaged CO absorption lines over the entire FoV, H band image, LOSV, and LOSV dispersion.}
    \label{fig:check_3_1566}
\end{figure}

The best model for NGC 1566 reveals recent star formation in both the NSC and the circumnuclear disk. The residual maps in Figures \ref{fig:k2_age_ak_NSC_1566} and \ref{fig:k2_age_ak_disk_1566} indicate that star formation in the disk occurred very recently or is ongoing, with minimum residuals for ages of less than $3\ \mathrm{Myr}$. The NSC also shows evidence of recent star formation, with ages younger than $10\ \mathrm{Myr}$, and a minimum of between $3$ and $10\ \mathrm{Myr}$. The extinction values are well constrained and higher compared to the other two objects, particularly in the NSC, which supports the proposed recent star formation in these structures.

The NSC in NGC 1566 is compact and relatively massive, and the disk has similar geometry to NGC 1433 with a size of $\sim 150\ \mathrm{pc}$, an aspect ratio of $(h/r)_{disk} \sim 0.25$, and a nearly face-on orientation ($i \lesssim 10\degree$). The central bright source of continuum emission is well represented by the model, but the depth of the CO absorption features is underestimated. This could be due to a large LOSV dispersion ($\sigma_v$), imprecision in determining the ages of the stars, or imperfections in the stellar templates. The observed LOSV map is well reproduced by the model, but the $\sigma_v$ map is more difficult to compare, as the observed map is mostly uniform. However, the model accurately represents the average value of the $\sigma_v$.

In order to obtain a better fit of the CO absorption depth, we tried to constrain the ages of the stellar population in the disk and/or in the NSC to be older than 10 Myr. This slightly increased the CO depth. However, the improvement was minor, and this constraint had a negative impact on the rest of the fit: the continuum was not well reproduced by the template, making the fitting procedure converge toward a solution with a very large NSC, replacing part of the disk. 




\section{Discussion}
\label{sec:disc}
\subsection{Age and extinction}

By fitting the low-resolution continuum and CO absorption features, we determined ages and extinction for the stellar populations of the two components solely based on the SINFONI observations. 

For NGC 1808, we find evidence of past star formation with young stellar populations of ages between 10 and 50 Myr. These estimates are in good agreement with previous results, which all point toward the presence of recent stellar populations in the central region: \citet{Kotilainen1996} also estimate the age of the stellar populations in the central region to be between $\sim 10$ and $\sim 40$ Myr from NIR line and continuum analysis, \citet{Galliano2005, Galliano2008} observed the presence of a very embedded young stellar cluster in the MIR, and \citet{Busch2017} measured the age of the 1 kpc circumnuclear ring to range between 5 and 8 Myr (the age of the central cluster is not given, but is indicated as having experienced less recent star formation).

For NGC 1433, we find no evidence of star formation in the last 100 Myr. This result is supported by previous publications on the object \citep{Cid1998, Sanchez2011, Smajic2014}, which do not detect any significant star formation or extinction outside the 1 kpc circumnuclear ring. 

For NGC 1566, we find evidence of very recent or ongoing star formation, both in the NSC and in the 150 pc ring. This result is not in strong agreement with previous publications, which tend to attribute a significant portion of the nuclear featureless continuum to AGN activity (accretion disk + hot dust) instead  of hot stars \citep{Smajic2015, dasilva2017}. This contribution from the AGN is supported by the observation of associated emission lines, and is not taken into account in our model, which probably overestimates the contribution from young stars. However, these latter two papers still point toward the presence of recent star formation: \citet{Smajic2015} measure the presence of significant star formation and supernova rates through $Br_\gamma$ and [Fe II] emission lines, and the stellar spectral synthesis done with the STARLIGHT software in \citet{dasilva2017} attributes a major part of the featureless continuum to a young stellar population. The AGN interpretation alone cannot explain the observations; in particular, the spatial extent of the continuum source, nondiluted nuclear CO absorption features, and the continuum in the circumnuclear disk. The stellar population interpretation can explain most of the SINFONI observation, but does not take into account observations at other wavelengths, which reliably identify an AGN component. The nucleus of NGC 1566 most probably hosts both phenomena, and the contribution of a young stellar population predicted by our model should be interpreted as an upper limit.

Overall, our method to determine the age of the young stellar populations in galactic centers provides good results. For NGC 1808 and NGC 1433, the ages obtained are fully consistent with observations and previous publications. For NGC 1566, the  depth of the CO absorption features is not well reproduced by our model, and the contribution of young stellar populations might be overestimated by our model because of the presence of the featureless AGN. 

\subsection{Supernova rates}

\begin{figure}[h]
    \centering
    \includegraphics[width=0.8\linewidth]{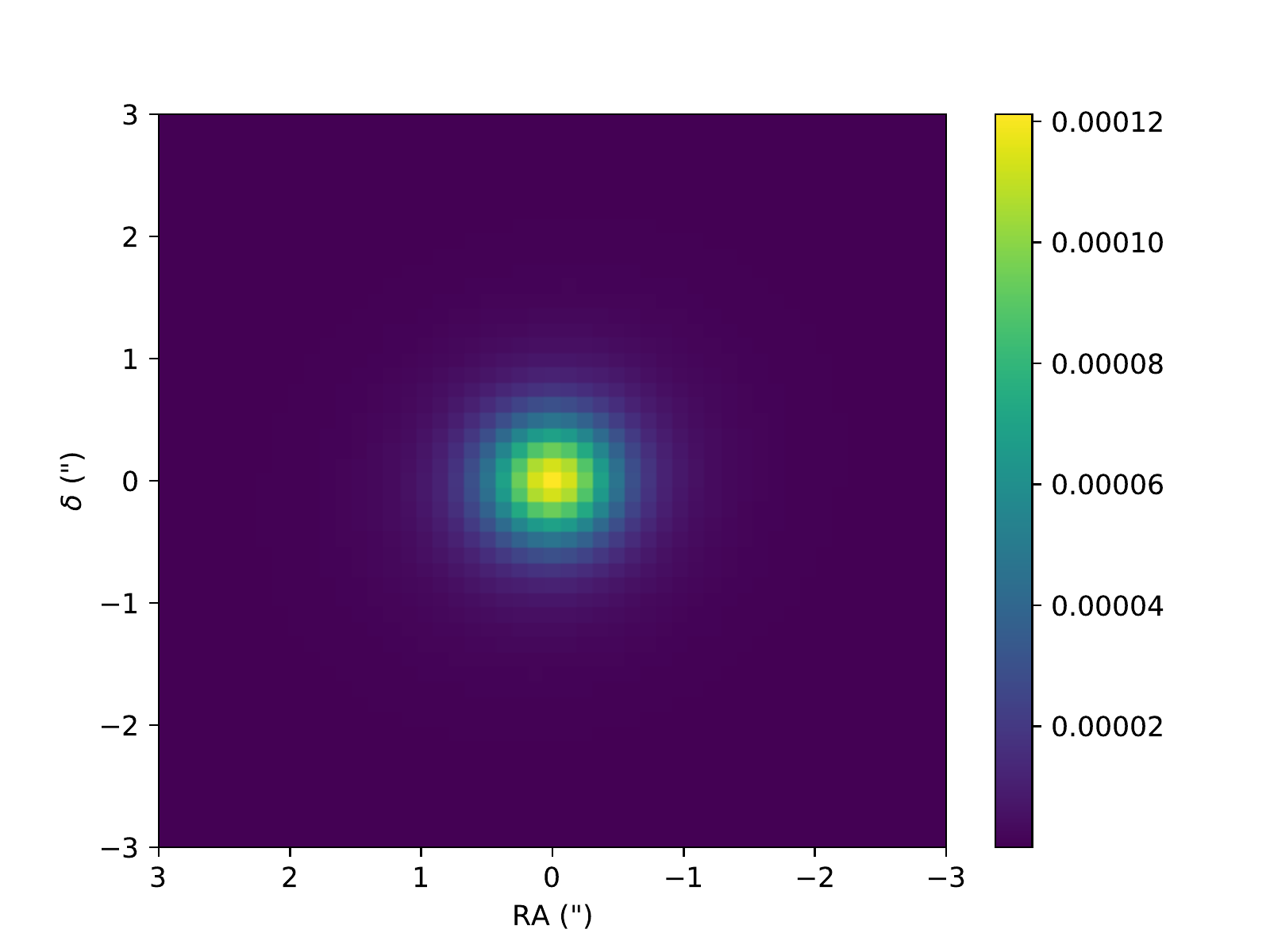}
    \includegraphics[width=0.8\linewidth]{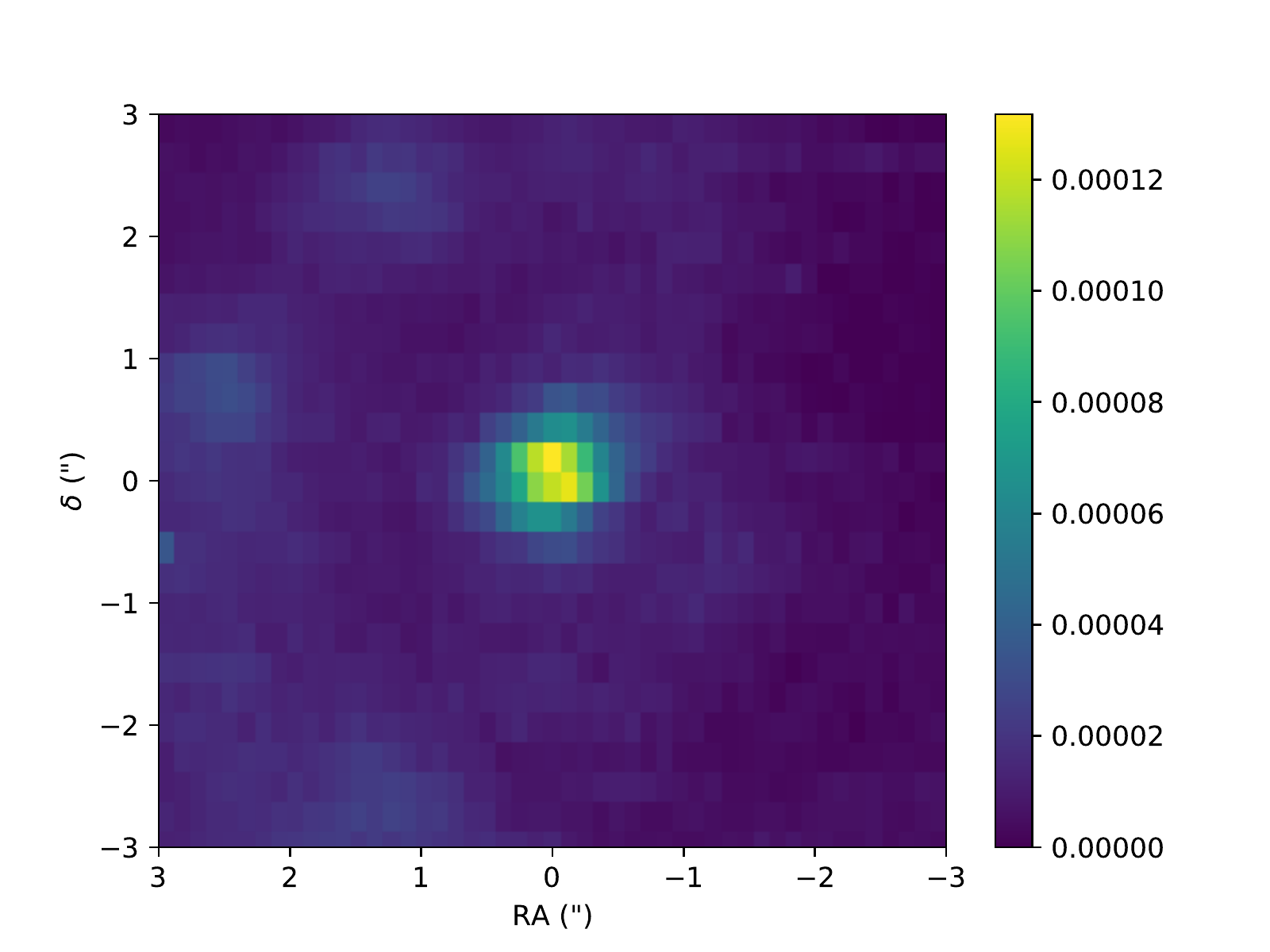}
    \caption{NGC 1808: 2D maps of the supernova rate (number of supernova per year and per pixel): (top) as estimated from our modeling and (bottom) as estimated from the [Fe II] emission line with the \citet{Rosenberg2012} conversion factor.}
    \label{fig:snr1808}
\end{figure}

\begin{figure}[h]
    \centering
    \includegraphics[width=\linewidth]{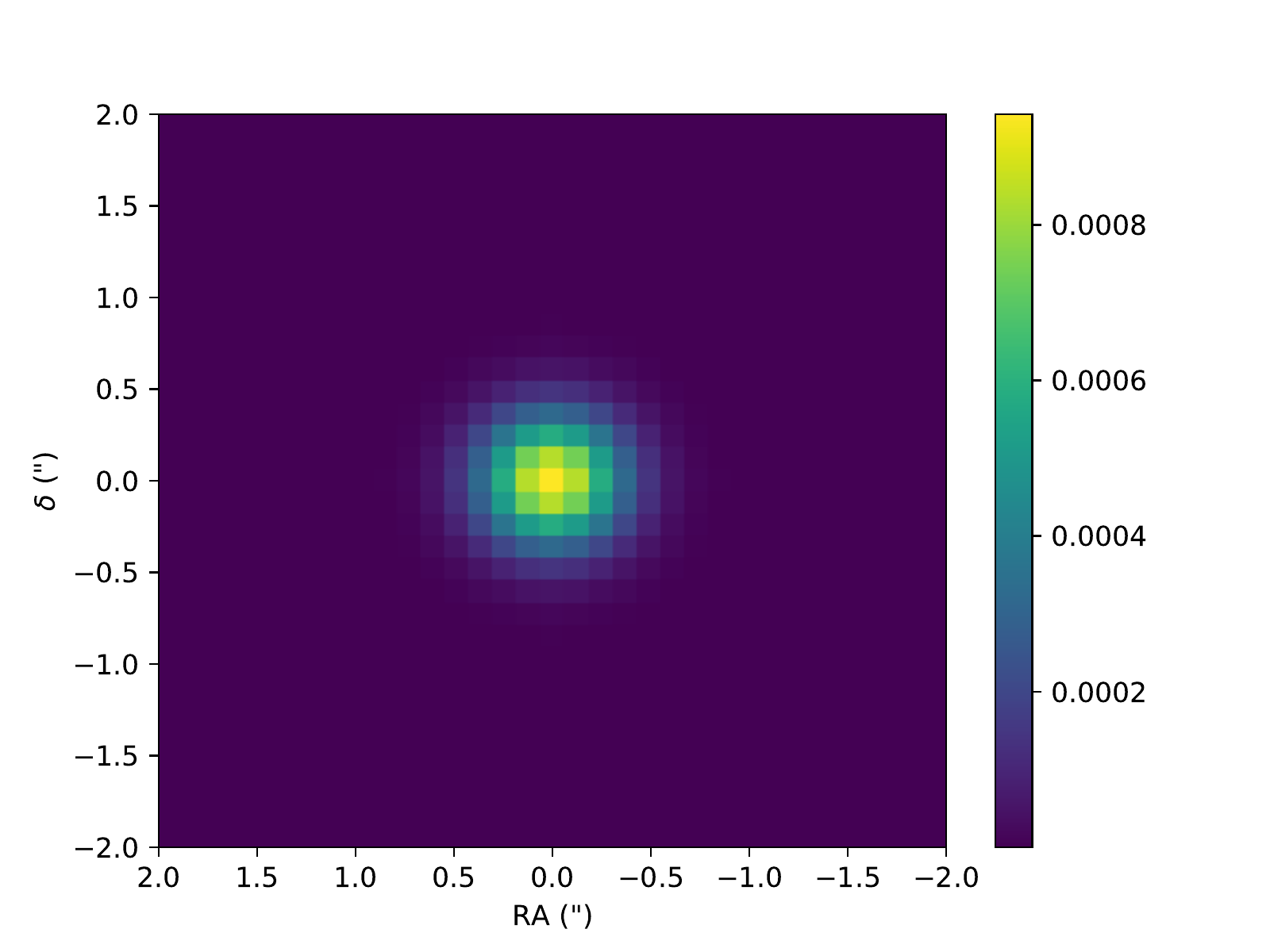}
    \includegraphics[width=\linewidth]{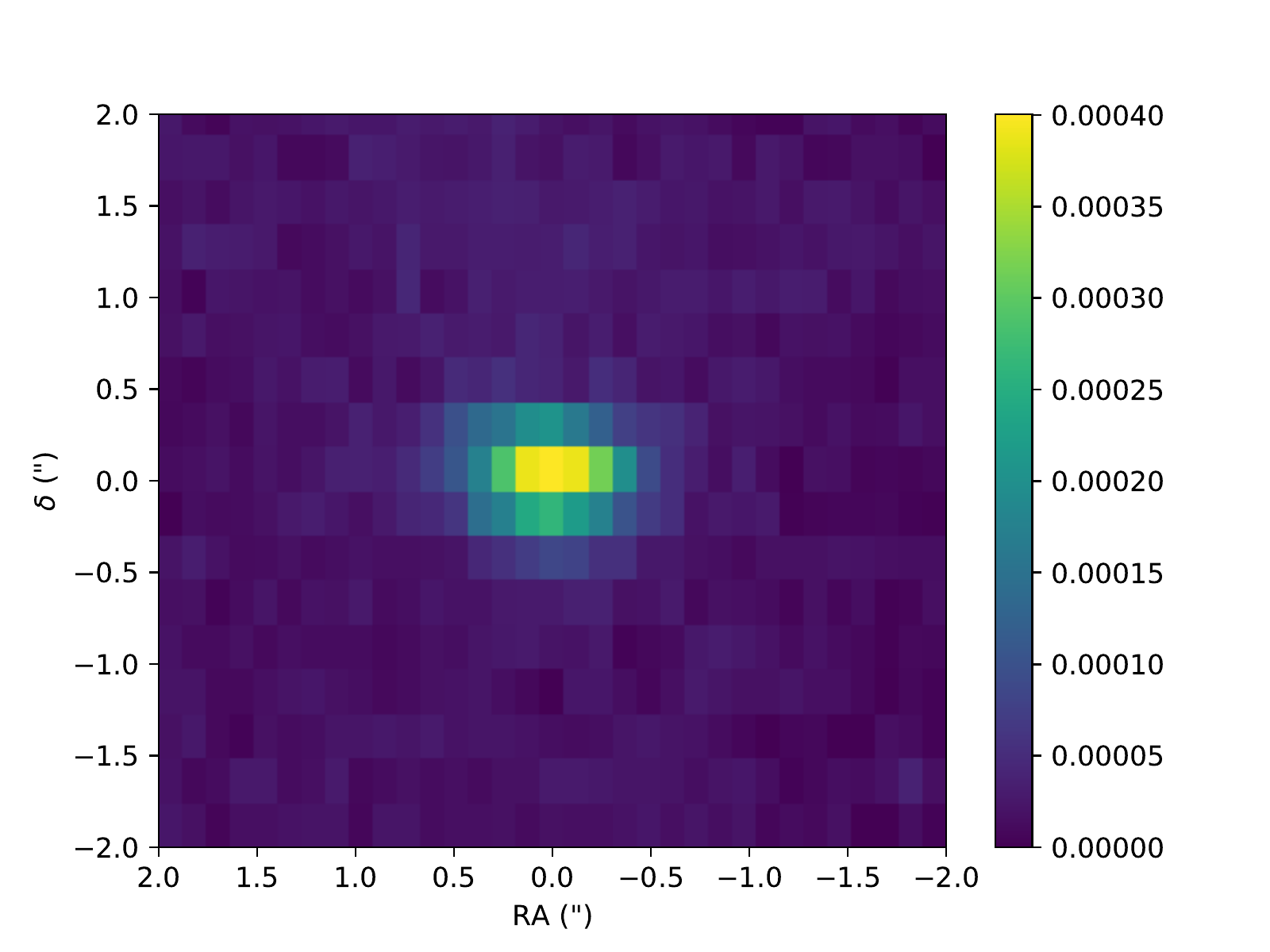}
    \caption{NGC 1566: 2D maps of the supernova rate (number of supernova per year and per pixel): (top) as estimated from our modeling and (bottom) as estimated from the [Fe II] emission line with the \citet{Rosenberg2012} conversion factor.}
    \label{fig:snr_1566}
\end{figure}

Figures \ref{fig:snr1808} and \ref{fig:snr_1566} show the supernova rate maps for NGC 1808 and NGC 1566, respectively, obtained from (1) our modeling; (2) the [Fe II] scaling relation from \citet{Rosenberg2012}. For NGC 1433, our model does not predict any supernova rate, and no [Fe II] emission line is detected.

For NGC 1808, the total supernova rate in the considered field of view amounts to 0.038, 0.028, and 0.007 supernov\ae\ per year with the three estimators. The relative distribution between the NSC and the disk is also well reproduced (see \ref{fig:check_3_11808}). The consistency between the estimate from our modeling and [Fe II] emission lines confirms the age and mass deduced from our model for the young stellar population. According to Fig. \ref{fig:snr_rate}, the stellar population is reaching the end of its supernova episode.

For NGC 1566, the results are very similar, but with a slightly larger discrepancy between our modeling and the [Fe II] estimators, in particular for the NSC where the difference between methods (1) and (3) is of an order of magnitude. However, even with the lowest estimator, the supernova rate is very significant in the central region, which supports the diagnostic of a young stellar population discussed above. According to Fig. \ref{fig:snr_rate} and our very young age estimates, this corresponds to the beginning of the supernova episode for this stellar population, and the high supernova rate should be sustained for several tens of millions of years.

The absence of [Fe II] emission in NGC 1433 is consistent with the prediction of our model, which does not detect any stellar population younger than 100 Myr, and so does not predict any core-collapse supernova.
Overall, the estimates of the supernova rate are of the same magnitude with the three methods. The consistency (same order of magnitude) between the estimate from our modeling and the direct [Fe II] measurements confirms the age determination we performed based on CO and continuum.

\subsection{Mass and geometry}

The simultaneous fit of the continuum of emission, LOSV, and $\sigma_v$ allows us to obtain an estimate of the 3D geometry of the object, in particular the inclination and scale height of the disk. For NGC 1808, the model converges toward a rather compact and very thick disk inclined by $\sim 50\degree$. The inclination found for the disk is fully consistent with values derived from the geometry of larger scale structures, such as the 1 kpc starburst ring or the de Vaucouleurs profile of the galaxy. With this inclination, the high aspect ratio of the disk ($r/h \sim 0.65$) is imposed by the LOSV to $\sigma_v$ maps ratio. The required mass to explain both LOSV and $\sigma_v$ with this inclination is consistent with the observed continuum emission flux given the assumed stellar populations. This consistent description of the four observables gives us confidence that such a compact thick disk is present around the NSC. Nevertheless, as already mentioned, a more standard larger and flatter disk is also probably present.

The geometry of our model for the disk of NGC 1433 and NGC 1566 consists of a larger and flatter disk seen almost face-on. The inclinations differ significantly from the estimates found from a  large-scale fit of the galaxies, which indicates inclinations of $\sim 30\degree$ for both objects from HI geometry and kinematics \citep[respectively]{Elagali2019, Ryder1996}. However, as noted in these latter papers, inclination is difficult to accurately measure so close to a face-on orientation, especially considering that strongly barred late-type galaxies could have an intrinsic elongation on average. Both results (from large-scale HI and small-scale stellar content) could be simultaneously correct in a tilted disk configuration.

\section{Conclusions}

\label{sec:con}
We used SINFONI spatially resolved spectroscopic observations of three nearby galactic nuclei (NGC 1433, NGC 1566, and NGC 1808) to derive the properties of the stellar populations in their central few hundred parsecs. Our main goal is to determine the age, mass, and 3D geometry of the nuclear star cluster (NSC) and the surrounding extended thick stellar disk.

To achieve this, we propose a method that uses single stellar population (SSP) spectra to determine the age of the stellar populations and construct synthetic observations for our model. We simultaneously fitted the spatially resolved line-of-sight-velocity (LOSV) and its dispersion ($\sigma_v$), as well as the low-spectral-resolution NIR continuum and high-spectral-resolution CO absorption features for each pixel of the datacube in order to overcome common degeneracies between key parameters.

We determined the ages of the various stellar populations by fitting the low-resolution continuum of emission and CO absorption features with UMISSP, an SSP library developed specifically for this purpose. We then determined the 3D mass distribution of the stellar populations with a two-component model composed of a Plummer NSC surrounded by a Miyamoto-Nagai disk, which produces synthetic continuum and CO datacubes, as well as LOSV and $\sigma_v$ maps, which we directly compared to observations.

For each object, we obtain a model that is consistent with all the SINFONI observables, and is in general agreement with previously published studies:
\begin{itemize}
    \item NGC 1808 experienced relatively recent star formation ---that is, around 10 Myr ago in the NSC and 30 Myr ago in a surrounding compact thick disk. A strong supernova rate is suggested by both the [Fe II] flux and our age modeling.
    \item NGC 1433 has not experienced recent star formation. It is composed of the most massive NSC with $M_{NSC} \sim 1.3 \times 10^9\ \mathrm{M_\odot}$ and is surrounded by a large and flat disk seen  almost face-on. No significant supernova rate is measured.
    \item NGC 1566 has recently experienced or is currently experiencing a starburst, both in the disk and the central NSC. Its geometry is similar to that of NGC 1433, with a massive NSC surrounded by a large and flat disk seen almost face-on. The nucleus is at the beginning of a long episode of supernova.
\end{itemize}

For the three objects, we find good agreement between the supernova rate estimated with our model and that estimated from [Fe II] scaling relations, confirming the age and mass deduced from our model for the young stellar population.
We conclude that, based on an IFU observation alone, our method can be used to characterize the properties of the stellar populations in the central few hundred parsecs of nearby galactic nuclei.

\begin{acknowledgements}
We thank the anonymous referee for his useful comments. This work was made possible by the support of the international collaboration in astronomy (ASU mobility) with the number $CZ.02.2.69/0.0/0.0/18_053/0016972$ and the institutional project RVO:67985815. ASU mobility is co-financed by the European Union. 
\end{acknowledgements}

%
%

\bibliographystyle{aa}
\bibliography{biblio}

\begin{appendix}
\section{ Age determination of stellar populations}

\begin{figure}[h]
    \centering
    \includegraphics[width=\linewidth]{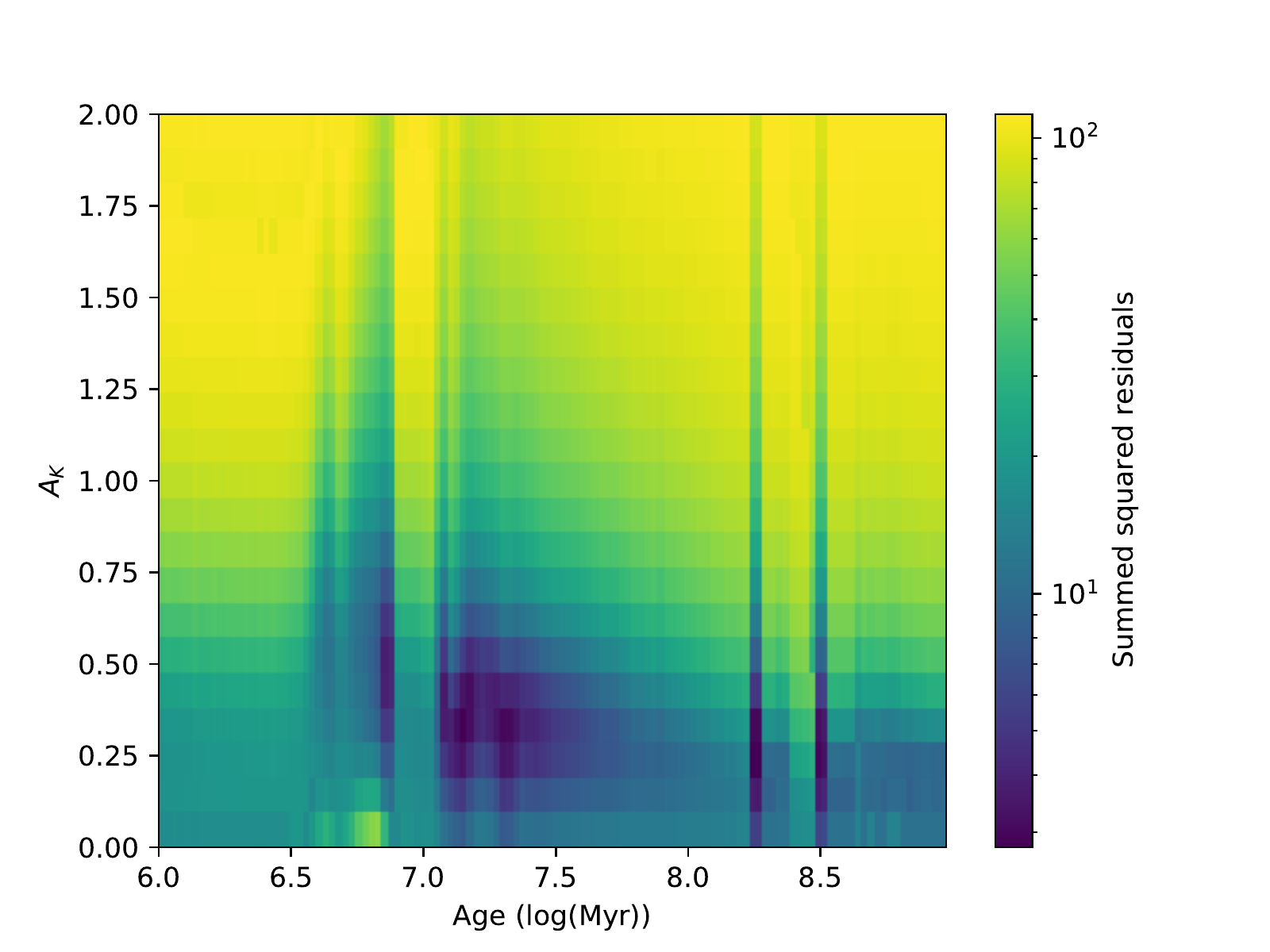}
    \includegraphics[width=\linewidth]{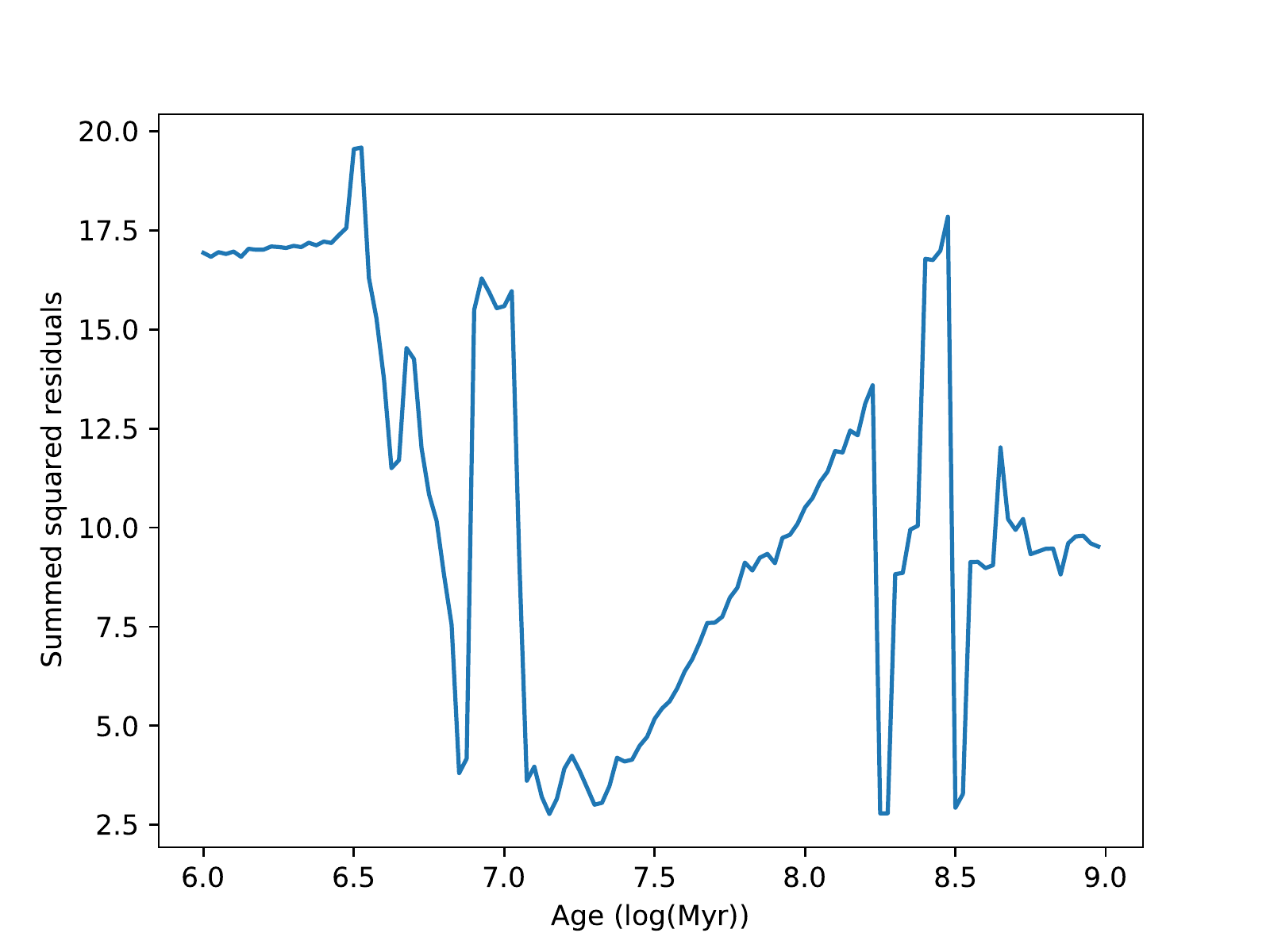}
    \includegraphics[width=\linewidth]{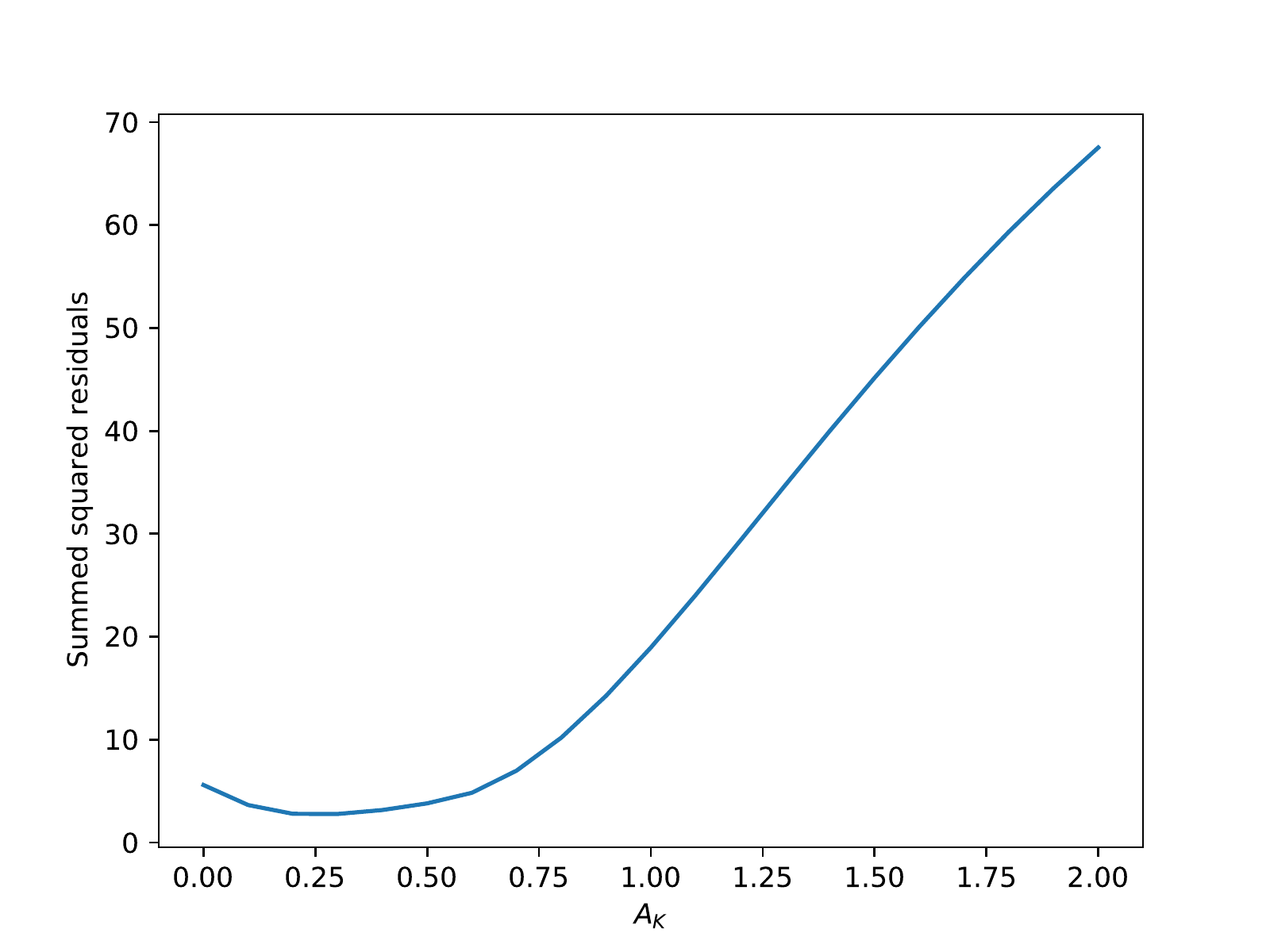}
    \caption{NGC 1808 (NSC): (top) 2D map of the  summed residuals as a function of age and extinction between the model and the extracted NSC spectrum, (middle) minimum residuals as a function of age, and (bottom) minimum residuals as a function of extinction.}
    \label{fig:k2_age_ak_NSC_1808}
\end{figure}

\begin{figure}[h]
    \centering
    \includegraphics[width=\linewidth]{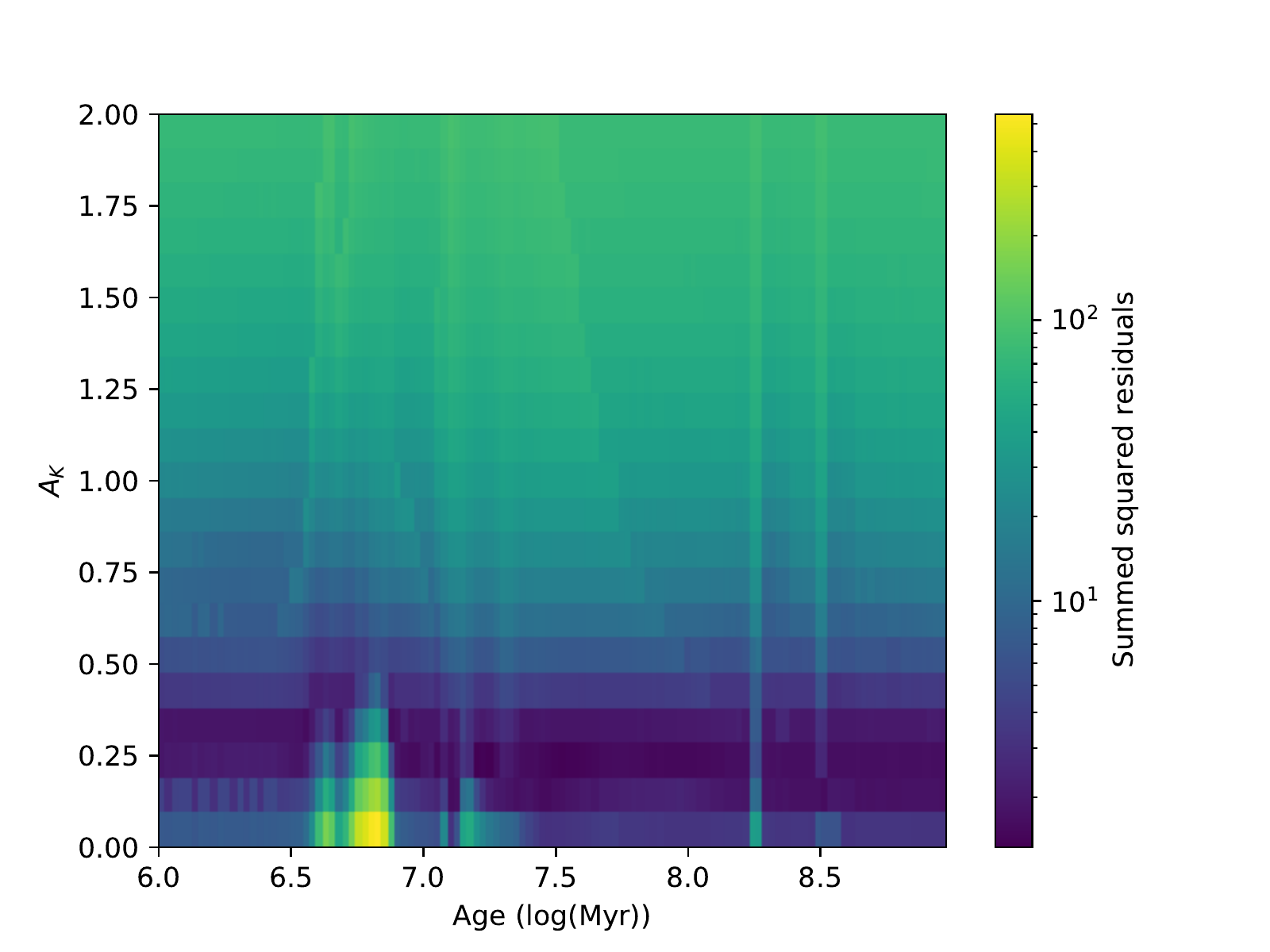}
    \includegraphics[width=\linewidth]{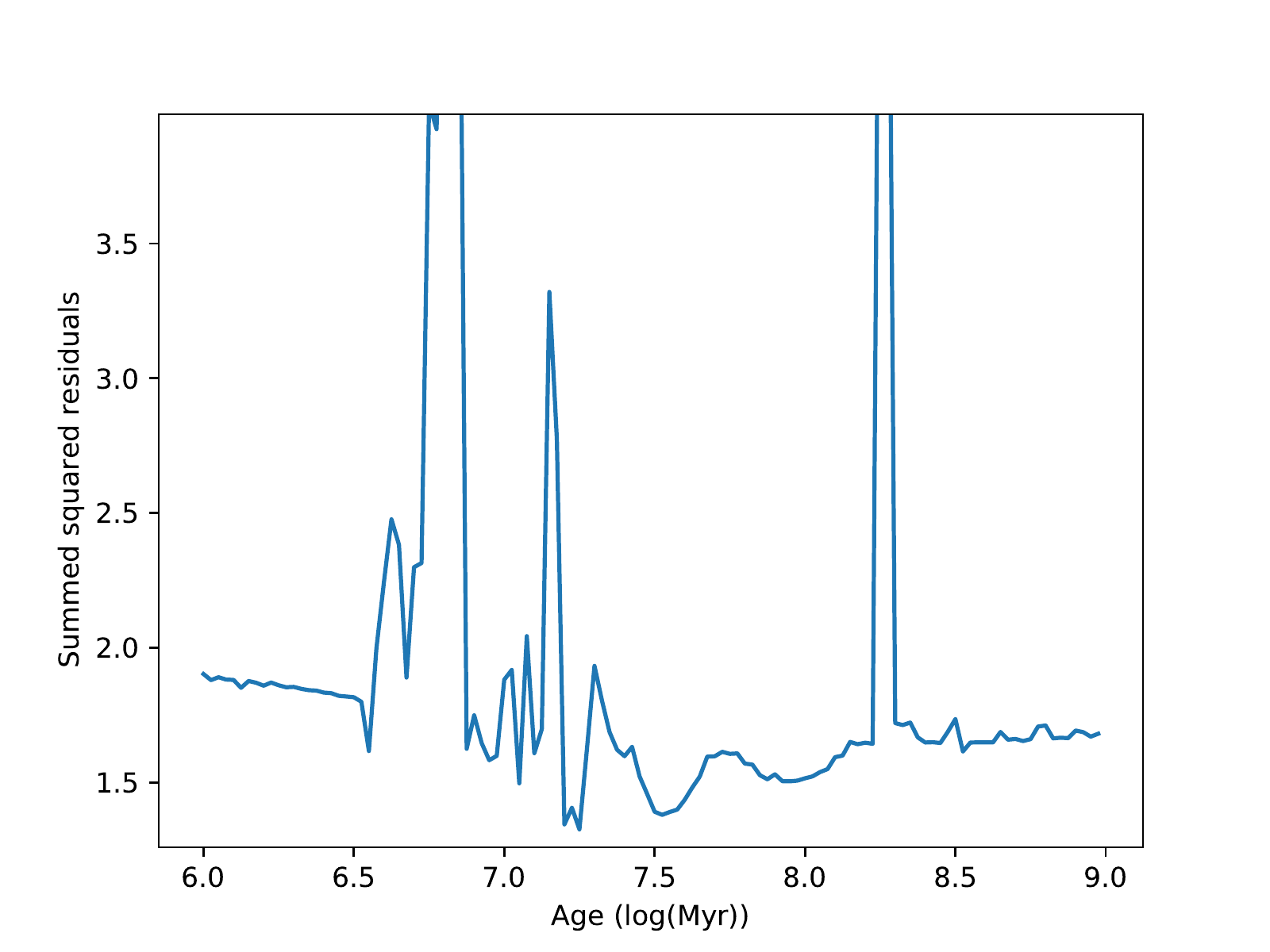}
    \includegraphics[width=\linewidth]{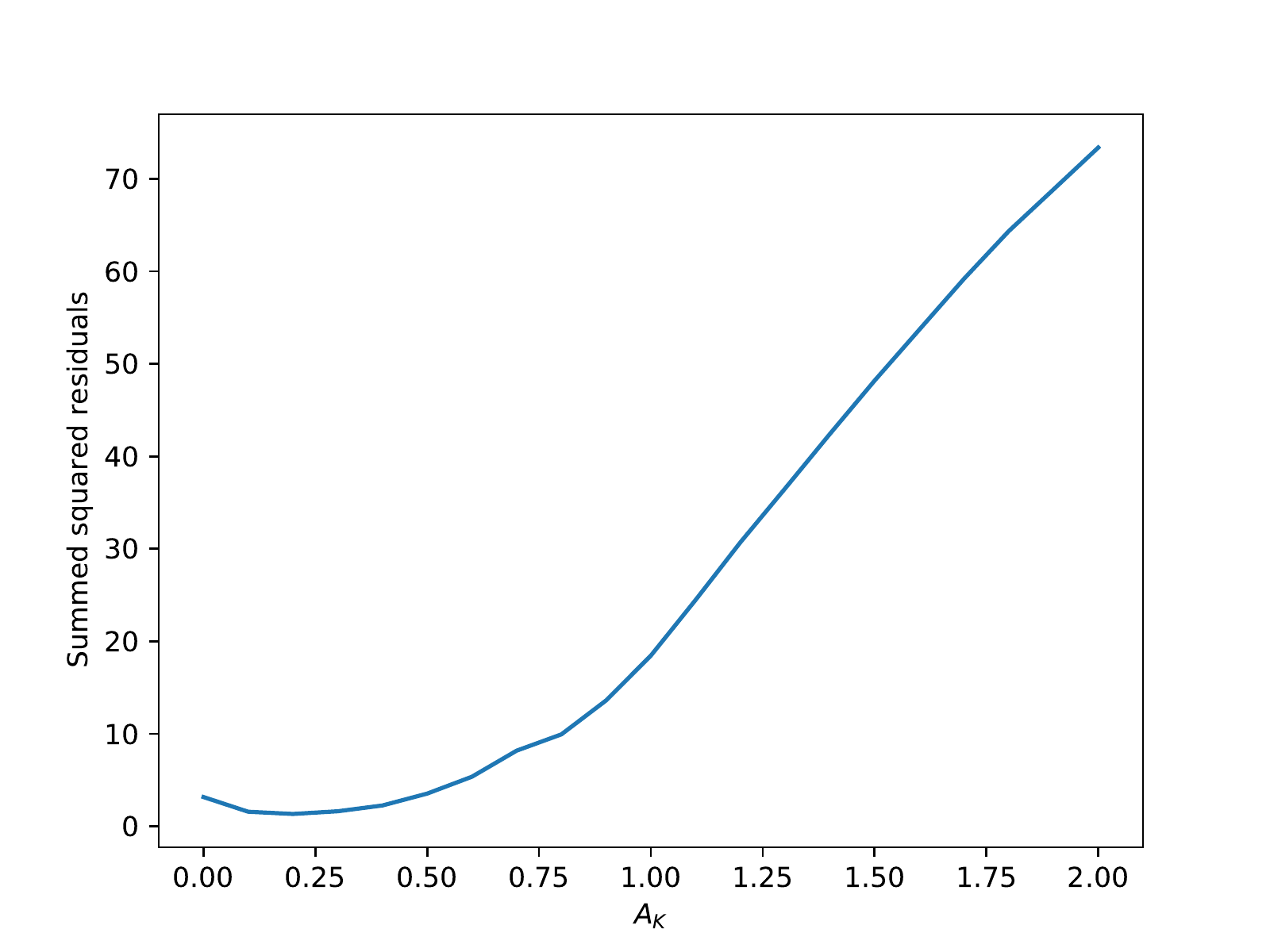}
    \caption{NGC 1808 (disk): (top) 2D map of the  summed residuals as a function of age and extinction between the model and the extracted disk spectrum, (middle) minimum residuals as a function of age, and (bottom) minimum residuals as a function of extinction}
    \label{fig:k2_age_ak_disk_1808}
\end{figure}

\begin{figure}[h]
    \centering
    \includegraphics[width=\linewidth]{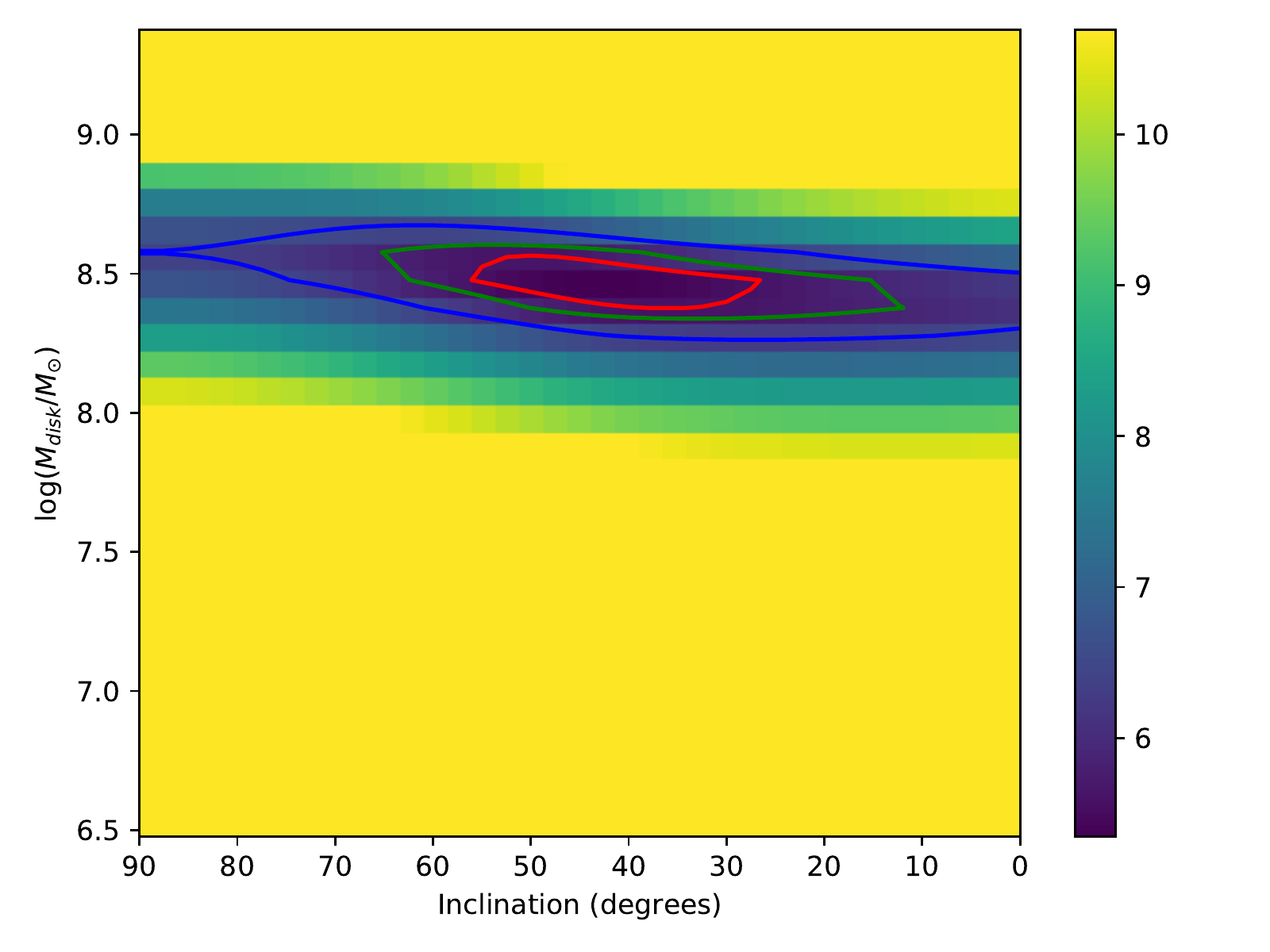}
    \caption{NGC 1808 (disk): 2D map of the summed residuals as a function of mass and inclination for the final model.}
    \label{fig:residuals_1808}
\end{figure}

\begin{figure}[h]
    \centering
    \includegraphics[width=\linewidth]{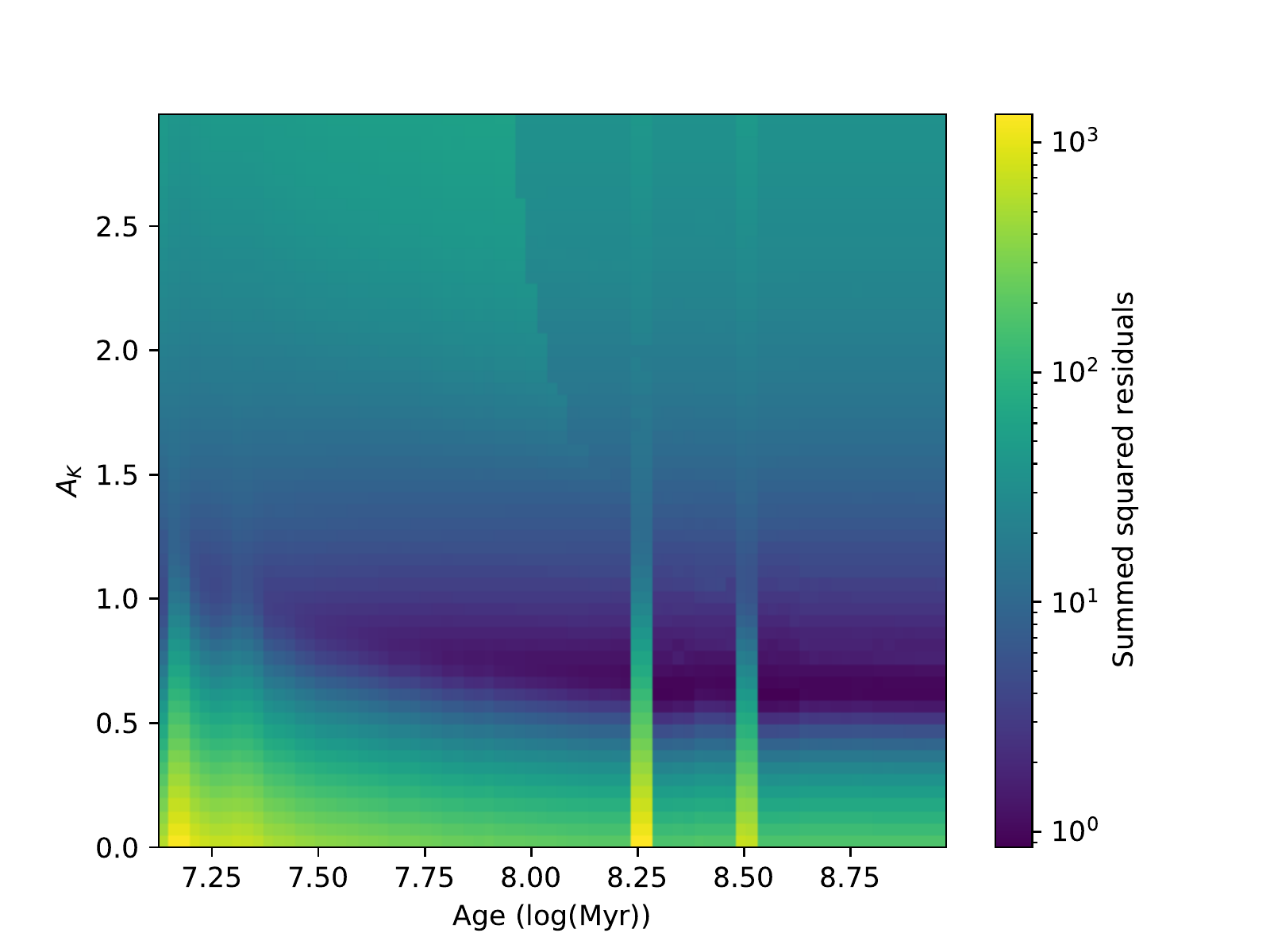}
    \includegraphics[width=\linewidth]{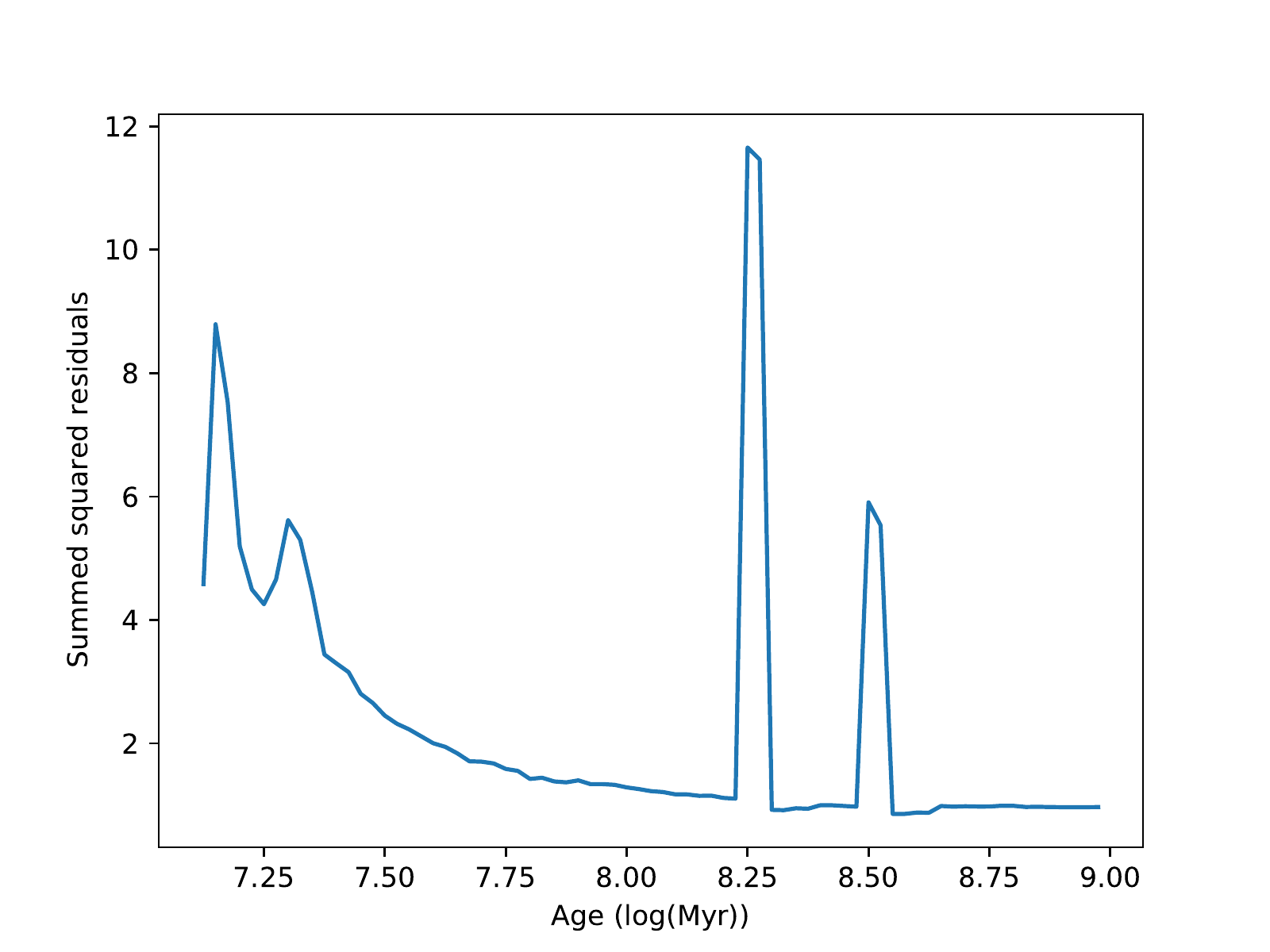}
    \includegraphics[width=\linewidth]{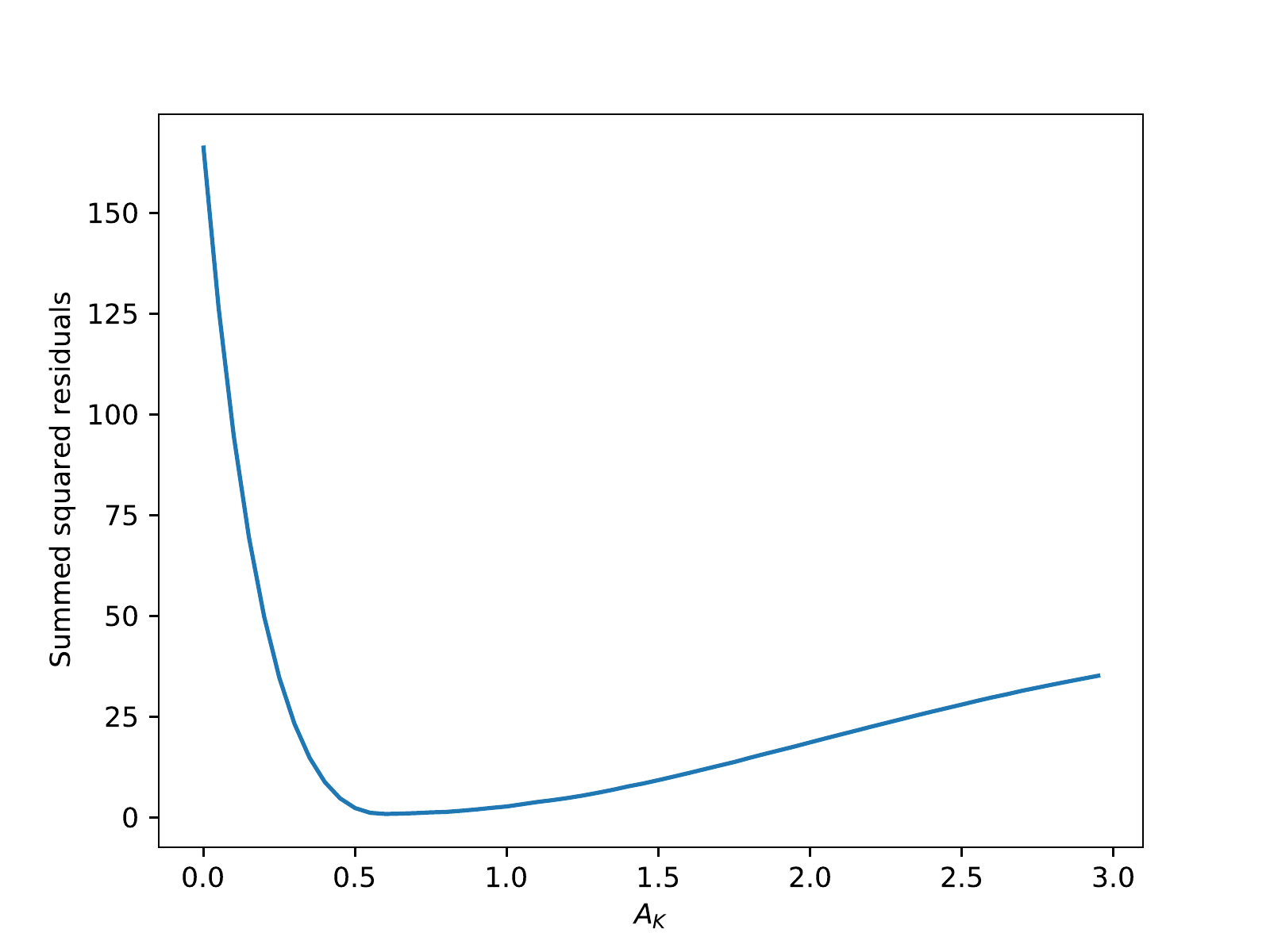}
    \caption{NGC 1433 (NSC): (top) 2D map of the summed residuals as a function of age and extinction between the model and the extracted NSC spectrum, (middle) minimum residuals as a function of age, (bottom) minimum residuals as a function of extinction.}
    \label{fig:k2_age_ak_NSC_1433}
\end{figure}

\begin{figure}[h]
    \centering
    \includegraphics[width=\linewidth]{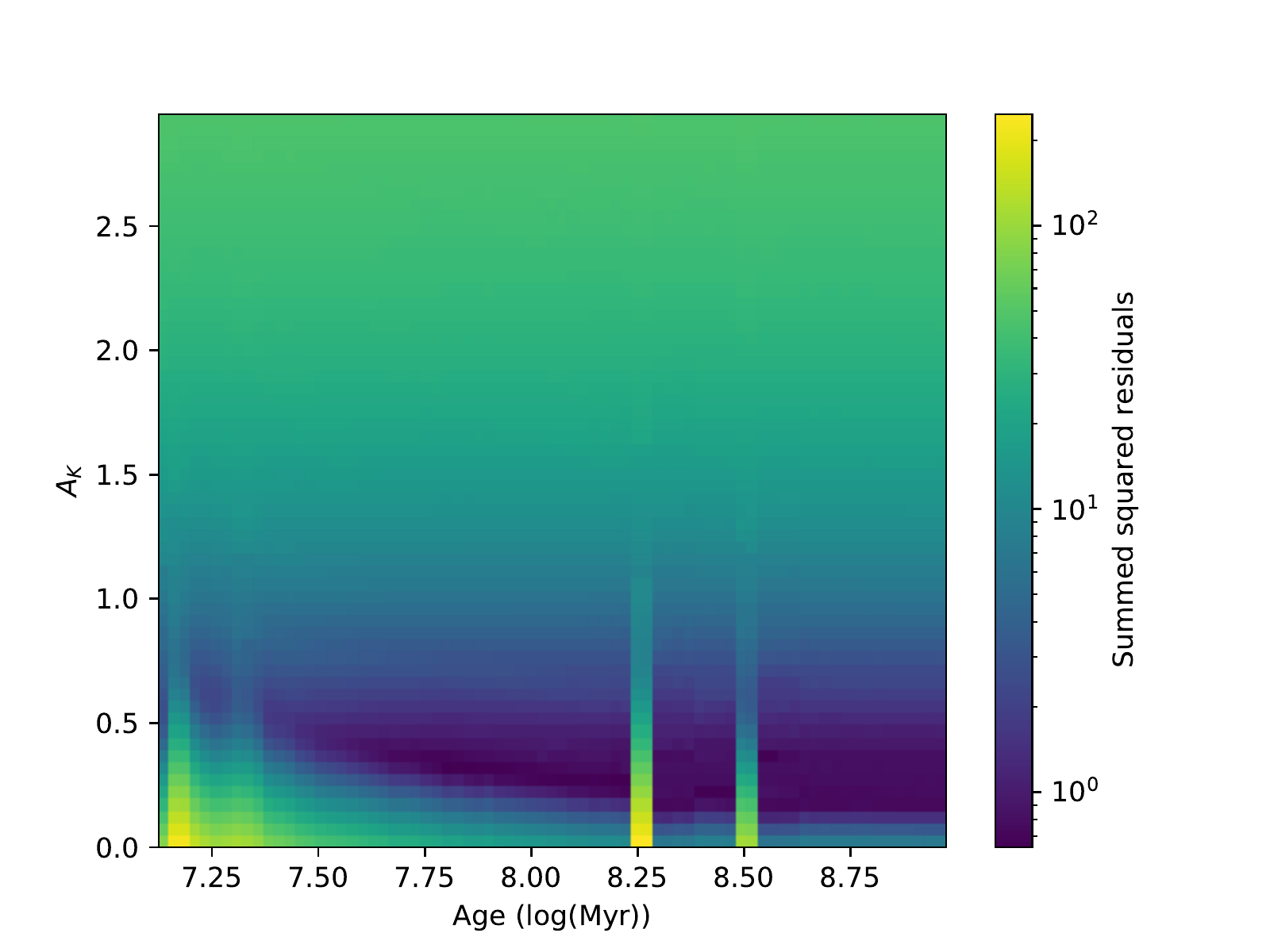}
    \includegraphics[width=\linewidth]{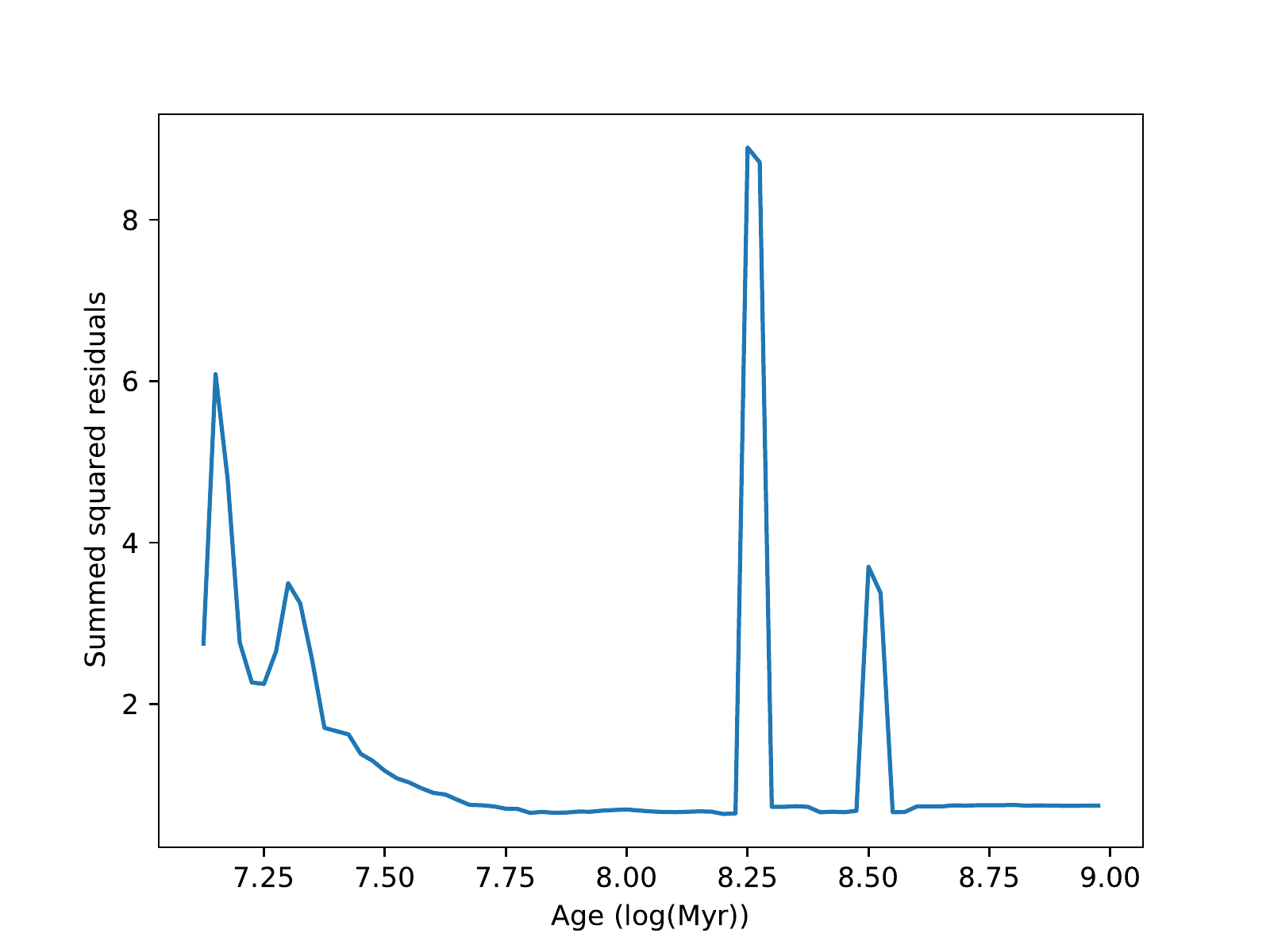}
    \includegraphics[width=\linewidth]{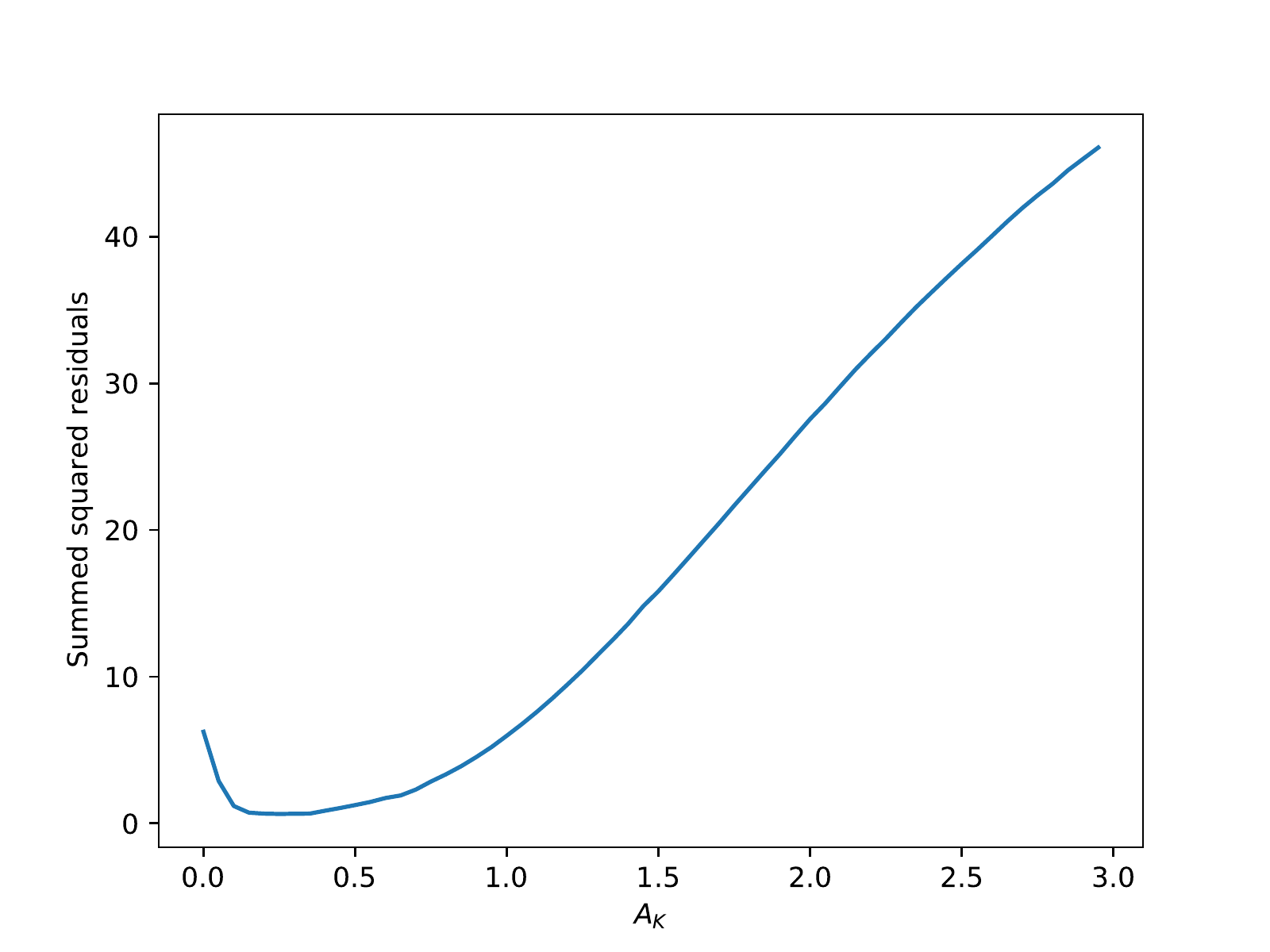}
    \caption{NGC 1433 (disk): (top) 2D map of the summed residuals as a function of age and extinction between the model and the extracted disk spectrum, (middle) minimum residuals as a function of age, and (bottom) minimum residuals as a function of extinction.}
    \label{fig:k2_age_ak_disk_1433}
\end{figure}

\begin{figure}[h]
    \centering
    \includegraphics[width=\linewidth]{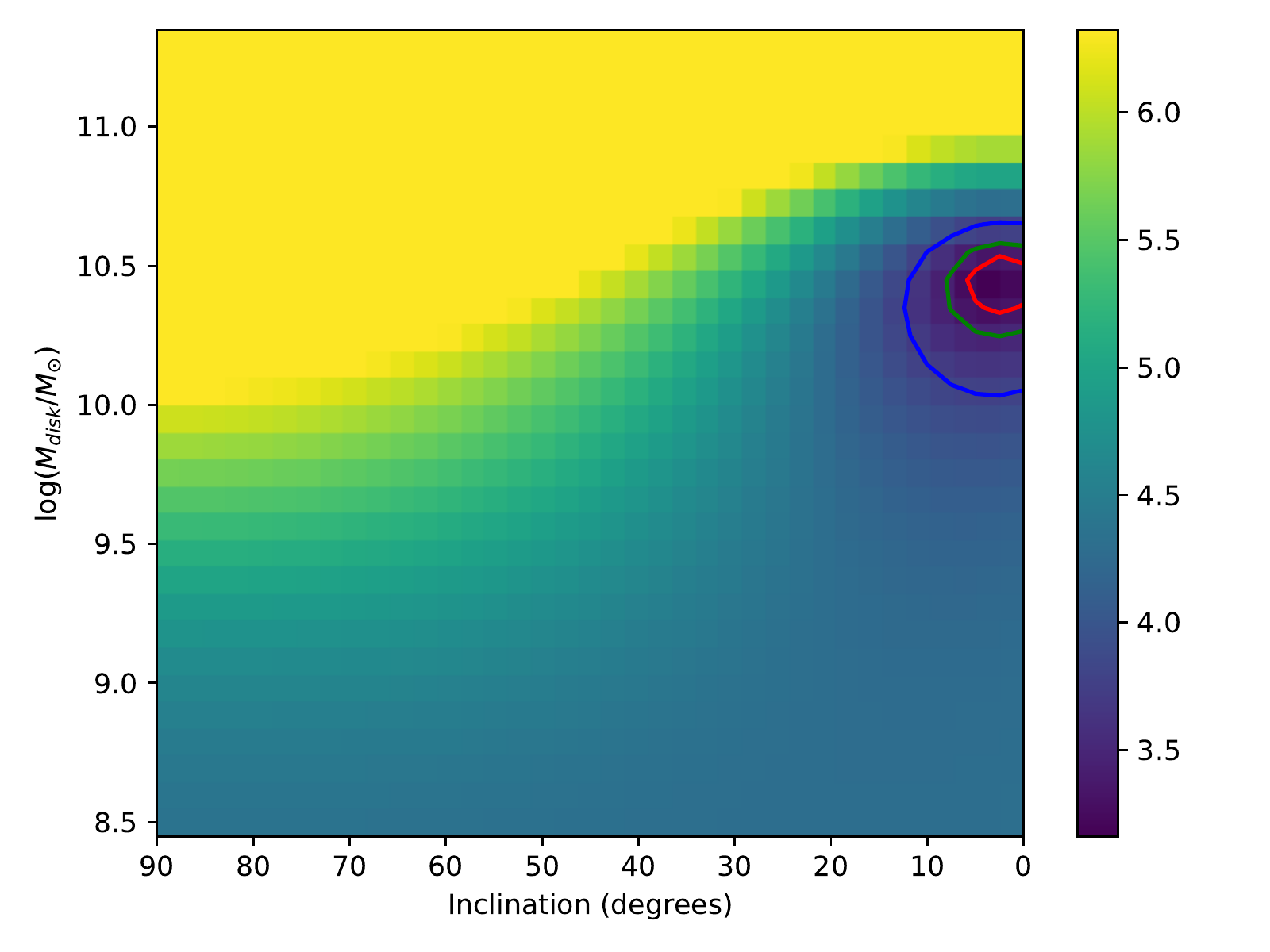}
    \caption{NGC 1433 (disk): 2D map of the summed residuals as a function of mass and inclination for the final model.}
    \label{fig:residuals_1433}
\end{figure}

\begin{figure}[h]
    \centering
    \includegraphics[width=\linewidth]{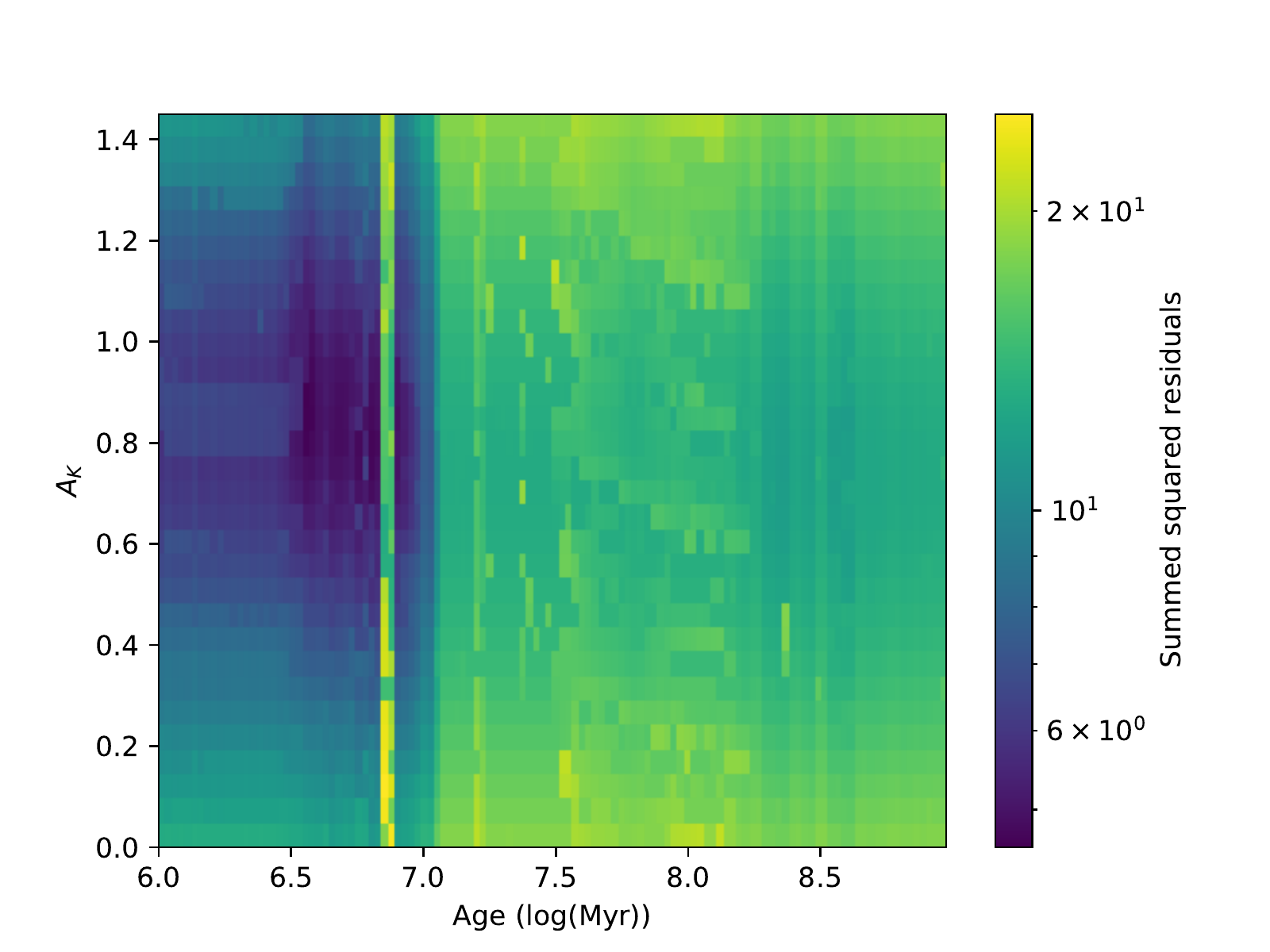}
    \includegraphics[width=\linewidth]{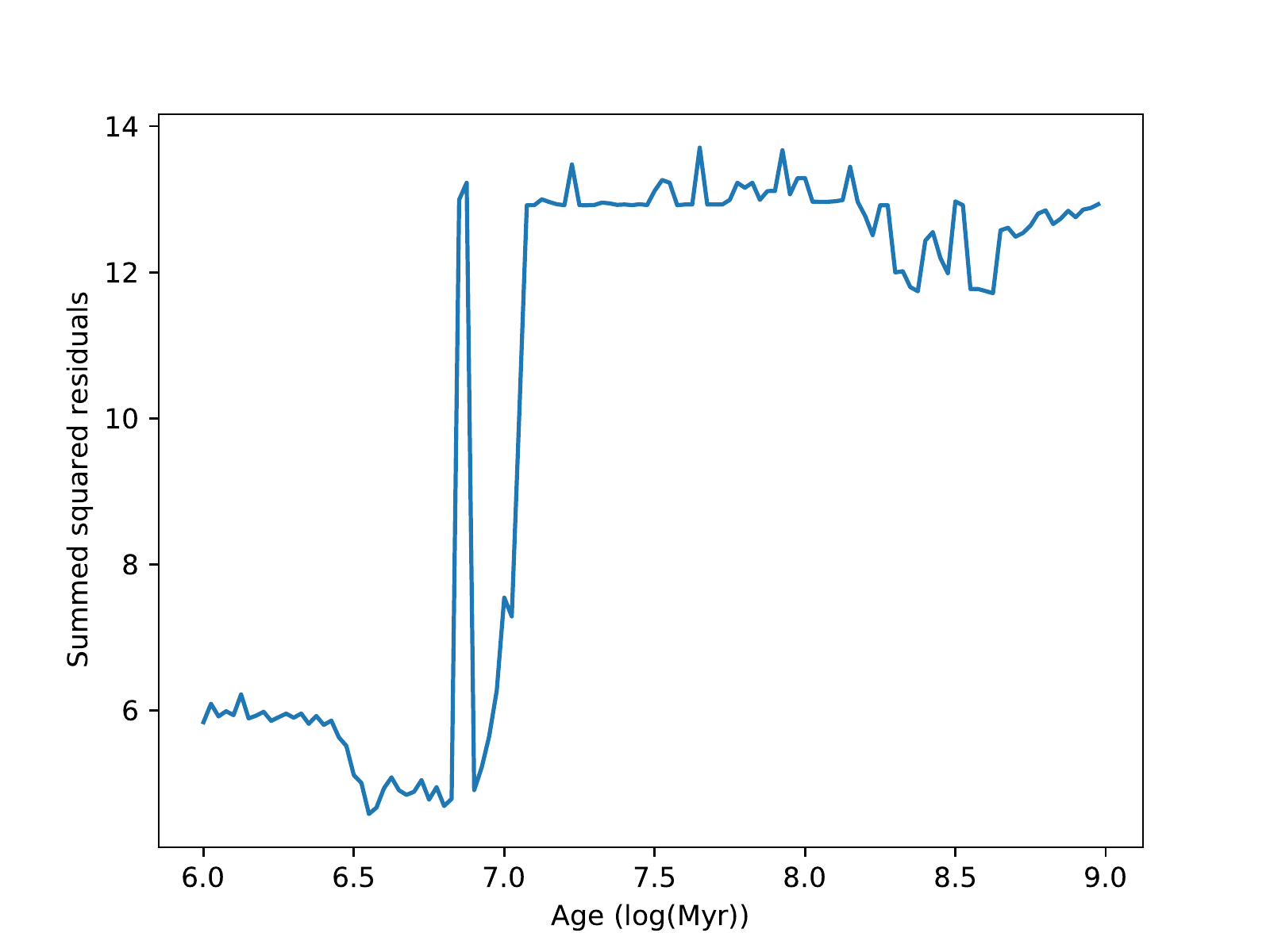}
    \includegraphics[width=\linewidth]{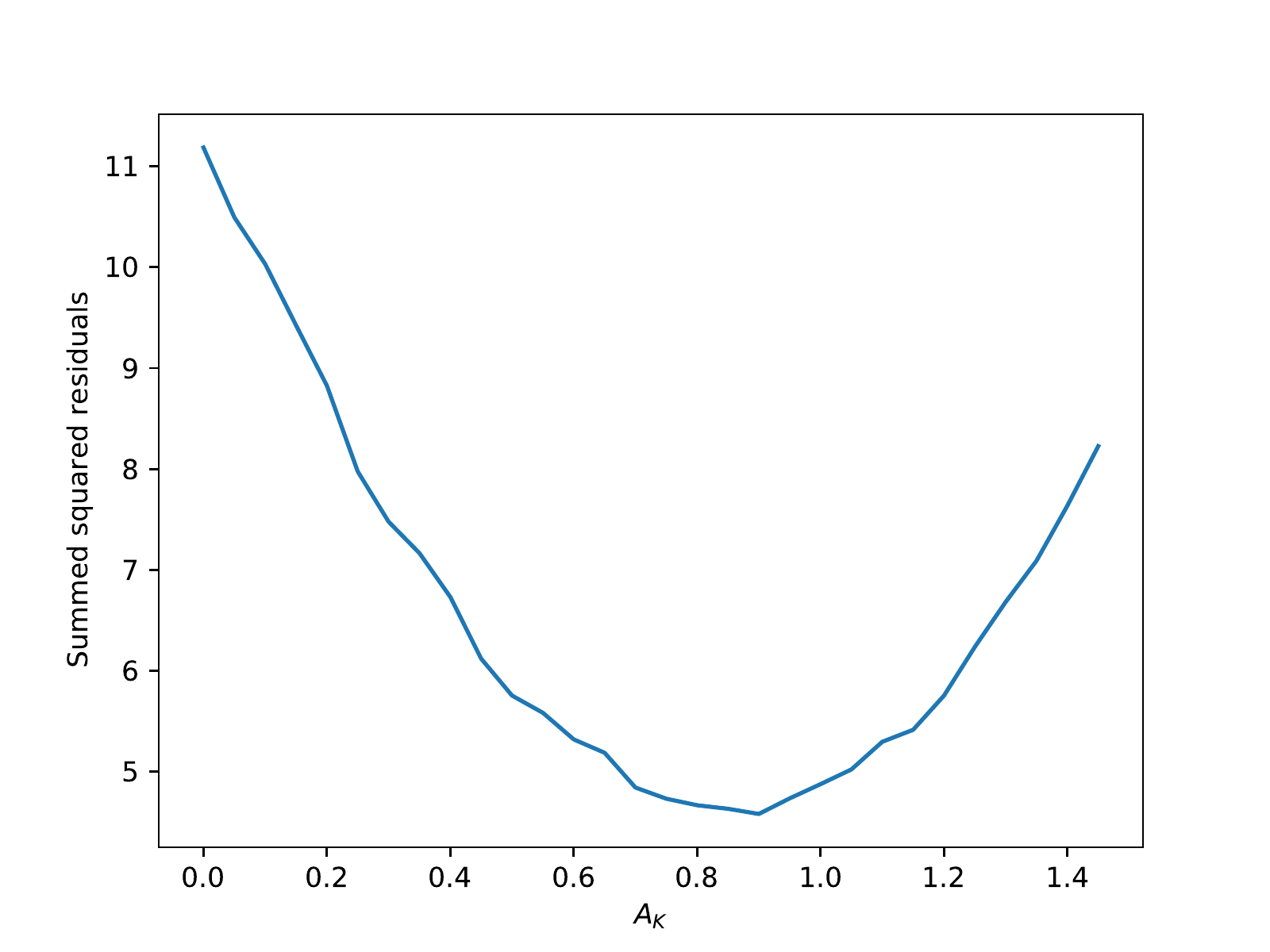}
    \caption{NGC 1566 (NSC): (top) 2D map of the summed residuals as a function of age and extinction between the model and the extracted NSC spectrum, (middle) minimum residuals as a function of age, and (bottom) minimum residuals as a function of extinction.}
    \label{fig:k2_age_ak_NSC_1566}
\end{figure}

\begin{figure}[h]
    \centering
    \includegraphics[width=\linewidth]{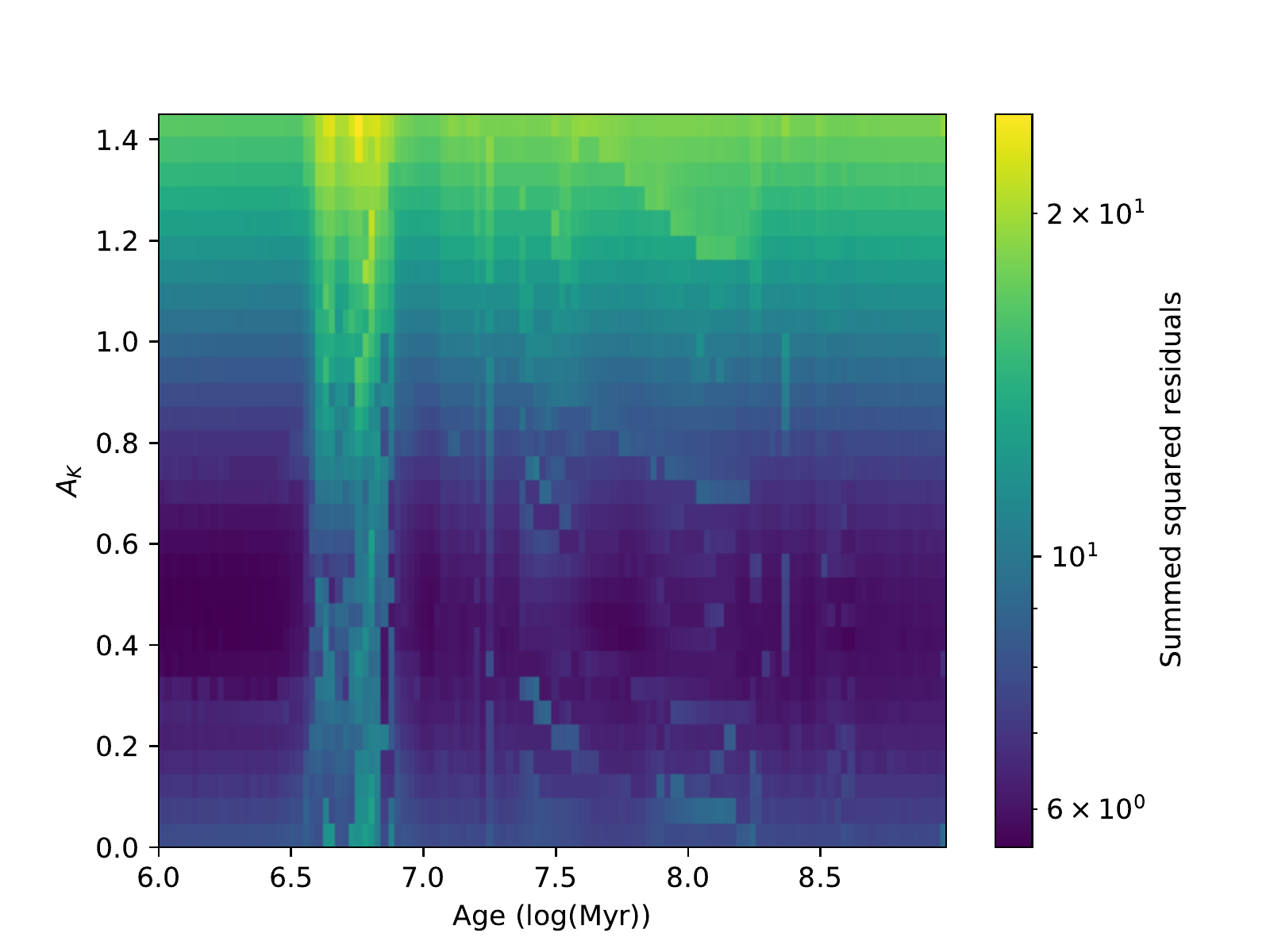}
    \includegraphics[width=\linewidth]{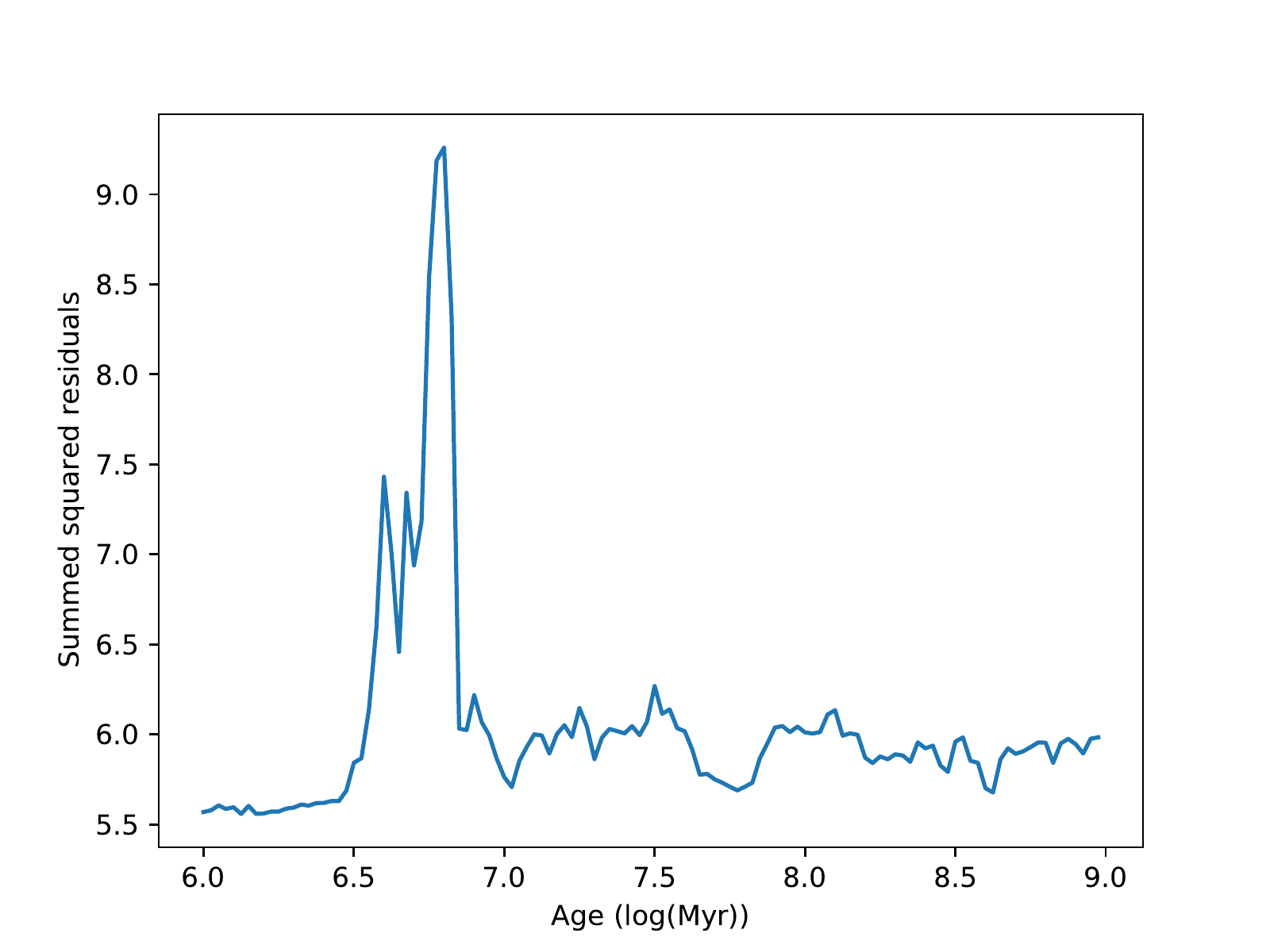}
    \includegraphics[width=\linewidth]{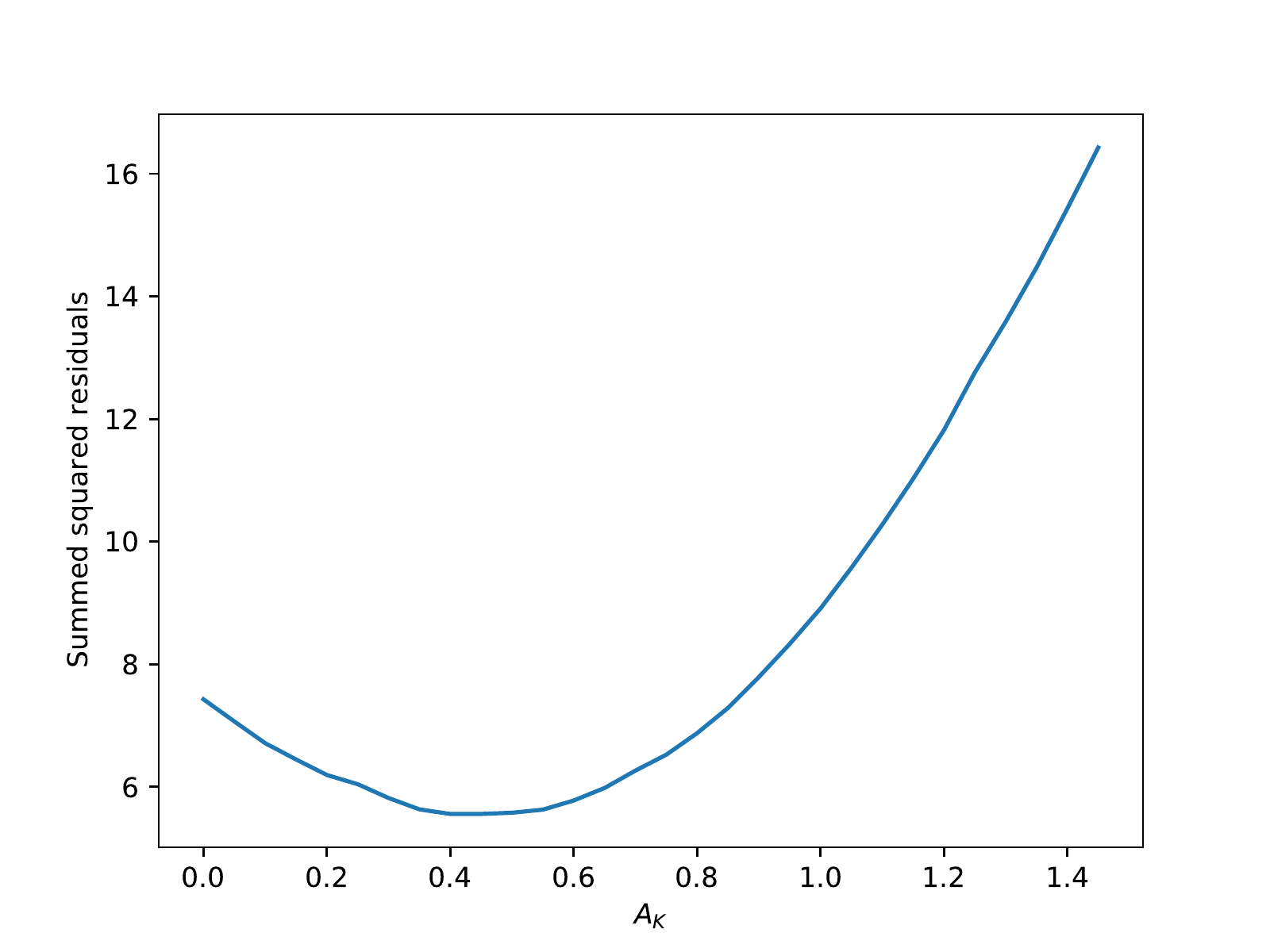}
    \caption{NGC 1566 (disk): (top) 2D map of the summed residuals as a function of age and extinction between the model and the extracted disk spectrum, (middle) minimum residuals as a function of age, and (bottom) minimum residuals as a function of extinction.}
    \label{fig:k2_age_ak_disk_1566}
\end{figure}

\begin{figure}[h]
    \centering
    \includegraphics[width=\linewidth]{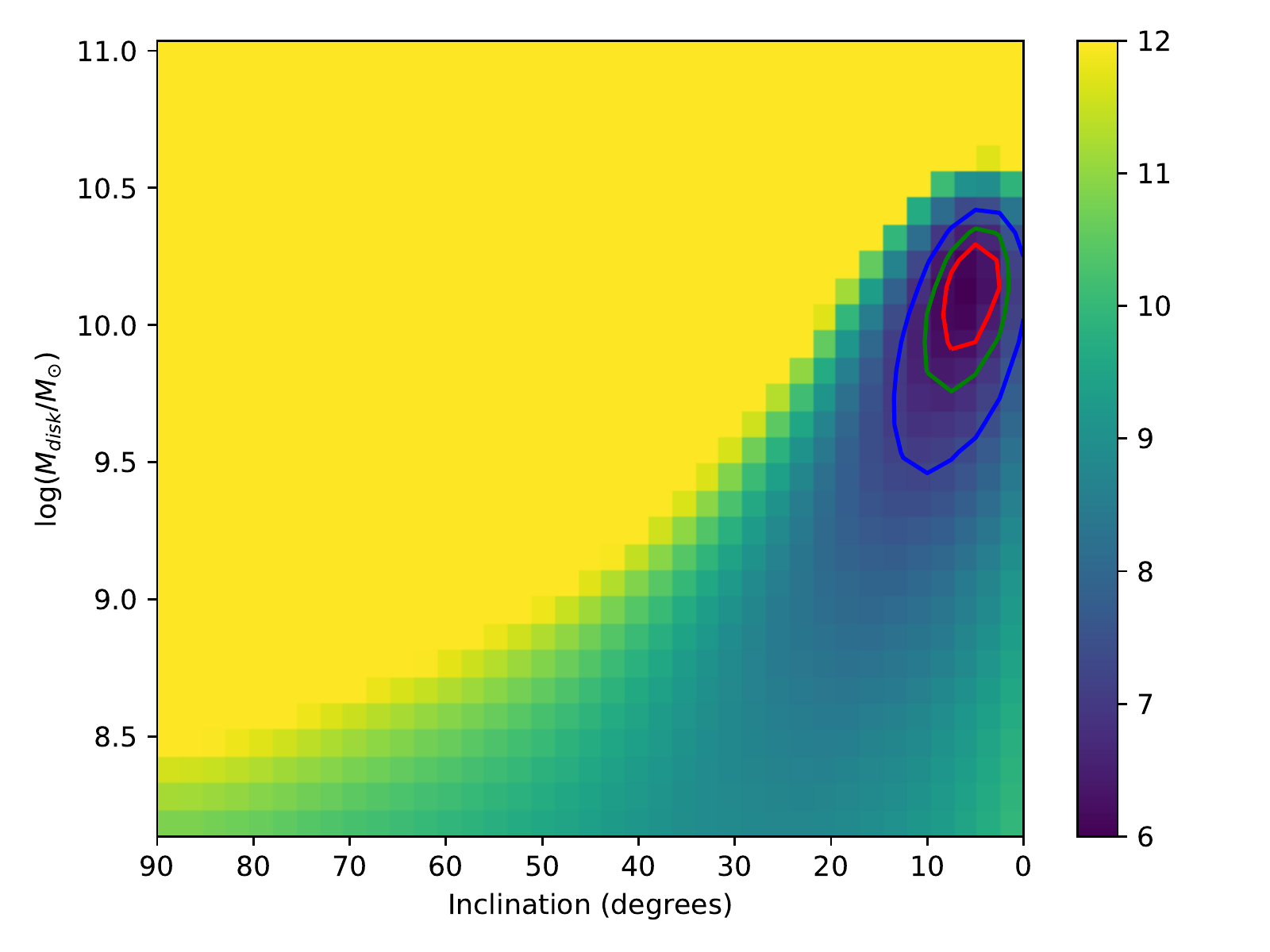}
    \caption{NGC 1566 (disk): 2D map of the summed residuals as a function of mass and inclination for the final model.}
    \label{fig:residuals_1566}
\end{figure}

In this Appendix, we present figures associated to the  procedure carried out to recover the ages  of the three objects: NGC 1808 (Figs. \ref{fig:k2_age_ak_NSC_1808}, \ref{fig:k2_age_ak_disk_1808} and \ref{fig:residuals_1808}), NGC 1433 (Figs. \ref{fig:k2_age_ak_NSC_1433}, \ref{fig:k2_age_ak_disk_1433} and \ref{fig:residuals_1433}), and NGC 1566 (Figs. \ref{fig:k2_age_ak_NSC_1566}, \ref{fig:k2_age_ak_disk_1566} and \ref{fig:residuals_1566}).

\newpage
\section{Comparisons}

\label{app:ssp}
The construction of SSP is a complex process where a great number of implicit astrophysical assumptions are made through the choice of the IMF, isochrones, and spectral library, leading to important variations between the output from one model and that of another. In order to check that no dramatic error has been made in the process, and to compare the typical age predictions of our models to others, we made a series of comparisons with three similar SSP models that also cover relatively young ages over a large spectral domain. The first one is GALEV \citep{Kotulla2009}, a user-friendly evolutionary model that produces low-resolution spectra over a wide range of ages similar to ours; the second is EMILES \citep{Vazdekis2016}, an SSP library with relatively high spectral resolution; the third is UNISSP, computed with the same isochrones as UMISSP, but using the PHOENIX v16 stellar spectra presented by \citet{Husser2013}. Please note that all the comparisons are done at solar metallicity, and thus do not include age--metallicity degeneracies.

\subsection{SED fitting}
\begin{figure}
    \centering
    \includegraphics[width=0.99\linewidth]{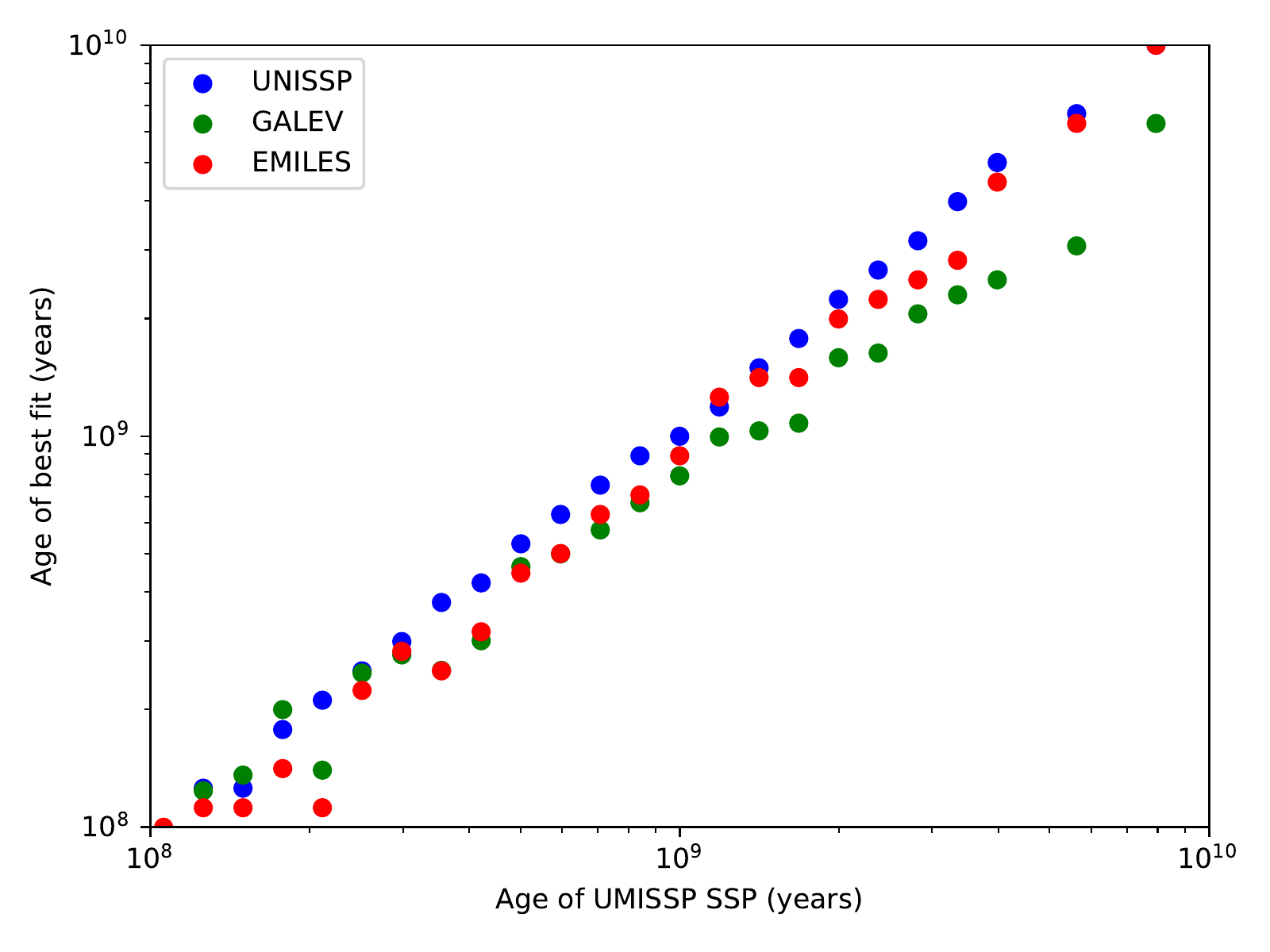}
    \caption{Retrieval of the age of UMISSP with the UNISSP, GALEV, and EMILES spectral libraries based on a SED fitting}
    \label{fig:comp_ages_sed}
\end{figure}

On the overlapping part of the parameter space, i.e., SSP from 100 Myr to 10 Gyr with solar metallicity, we extracted low-resolution SEDs from UMISSP at various ages and tried to retrieve the original age by fitting the spectra with the three other SSPs: UNISSP, EMILES, and GALEV. The results are presented in Fig. \ref{fig:comp_ages_sed}. While a visible difference is observed between the original and the recovered ages, we consider the result of this test to be conclusive, as the retrieved ages are systematically in the correct order of magnitude with a moderate spread. The largest discrepancies are observed for the youngest ages of UMISSPR (between 100 and 300 Myr), in particular when retrieved with the EMILES library. However, all models recognize a young stellar population, and these ages are outside the explicitly safe range of the EMILES SSP (which can be found on the EMILES website, and result from various limitations of the BaSTI and Padova00 models).

\subsection{NIR absorption line fitting}
\begin{figure}
    \centering
    \includegraphics[width=0.99\linewidth]{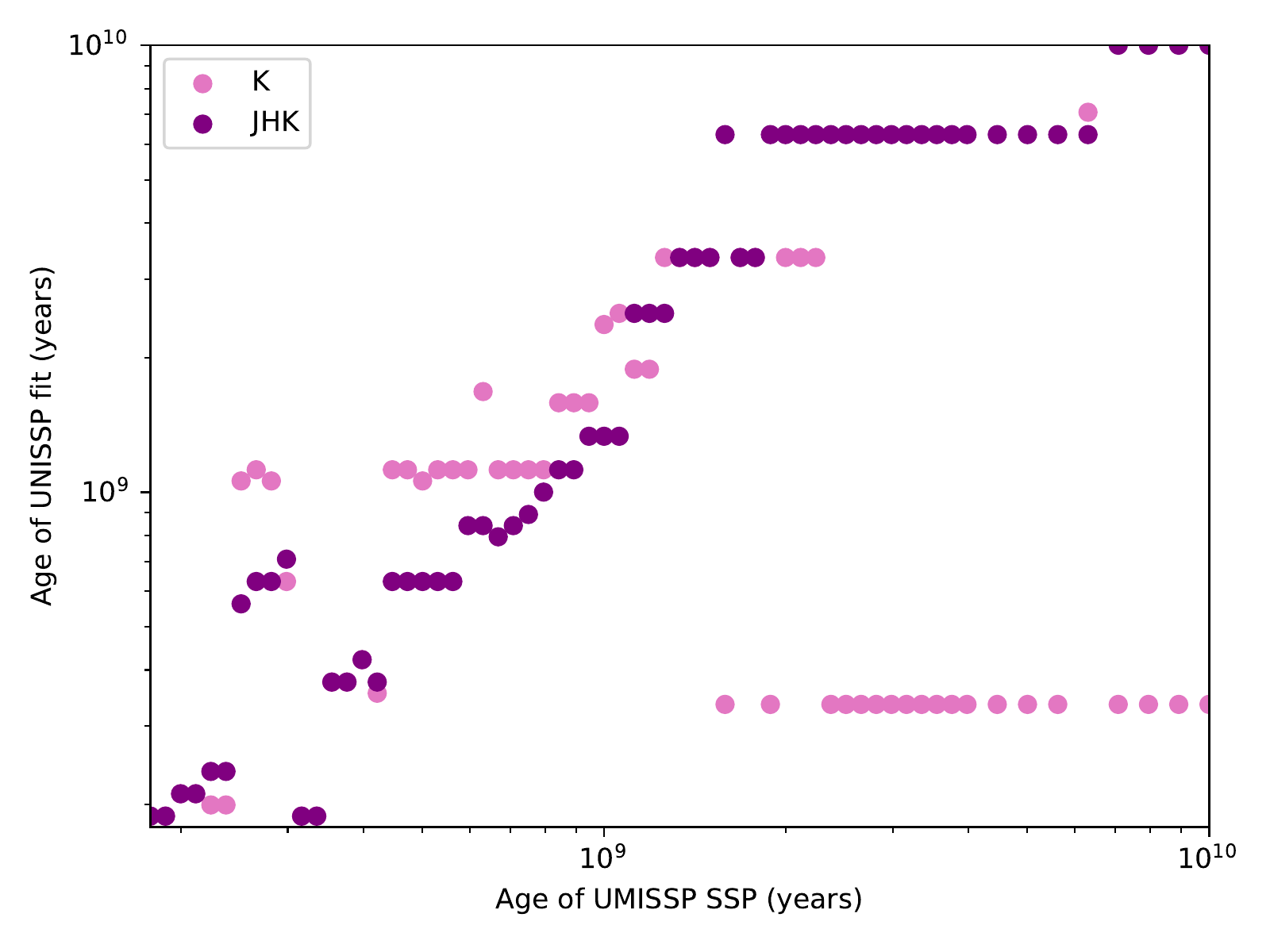}
    \caption{Retrieval of the age of UMISSP with the UNISSP library, solely based on the normalized absorption features between 1 and 2.5 $\mathrm{\mu m}$.}
    \label{fig:comp_ages_abs}
\end{figure}

Similarly, we checked the consistency of our two SSPs by retrieving the age of UMISSP solely by fitting its NIR absorption features with UNISSP. For this purpose, we extracted and normalized the spectra in two spectral domains (K band between 2.1 and 2.5 $\mathrm{\mu m}$, and JHK bands between 1. and 2.5 $\mathrm{\mu m}$), convolved them with a 1 nm Gaussian, and added a 5\% normal noise to simulate a $R\sim2000$ spectroscopic observation. The results are presented in Fig. \ref{fig:comp_ages_abs}. While the fit of the K band absorption features can produce good results in some cases, it can also lead to very large errors in the estimate of the age of a stellar population. Indeed, the CO bandheads, which arise from cold evolved stars and dominate the K band absorption spectrum, are remarkably similar over a wide range of ages. On the other hand, when simultaneously fitting the absorption features from the three NIR spectral bands (J, H and K), we are able to reliably retrieve the age of the original population. However, this procedure remains less precise than the SED fitting, and we suggest to couple the two strategies when fitting the age of an unresolved stellar population.

\subsection{Visible magnitude fitting}

We then checked the ability of our libraries to recover the ages of star clusters based on visible photometry, and compared the results with GALEV and EMILES. For this purpose, we used the M33 star cluster catalog presented in \citet{Fan2014}, which provides UBVRI photometry and age estimates for 188 clusters.

\begin{figure}
    \centering
    \includegraphics[width=0.99\linewidth]{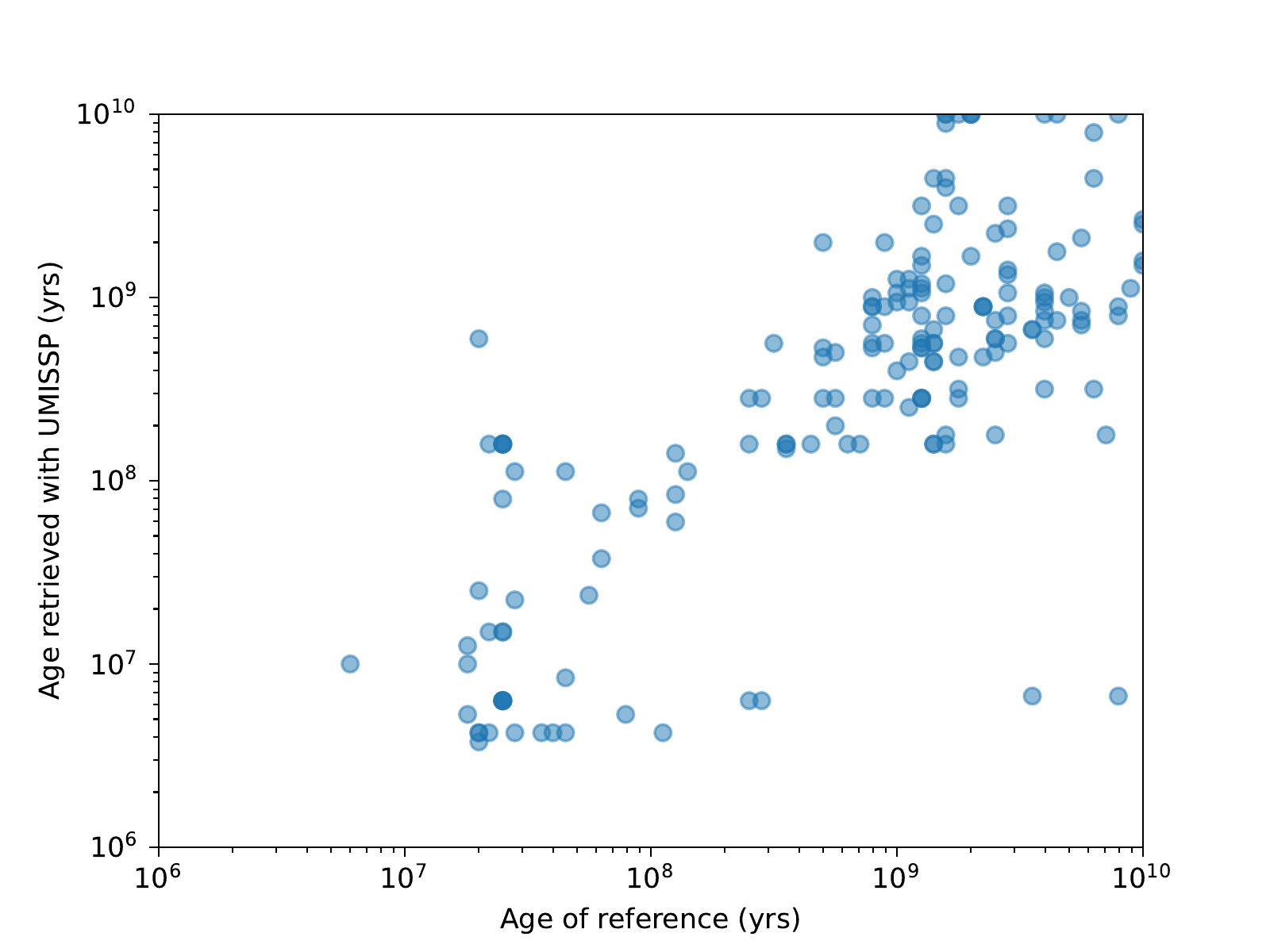}
    \caption{Retrieval of the M33 cluster age \citep{Fan2014} with the UMISSP library based on UBVRI photometry.}
    \label{fig:UMISSP_phot}
\end{figure}

\begin{figure}
    \centering
    \includegraphics[width=0.99\linewidth]{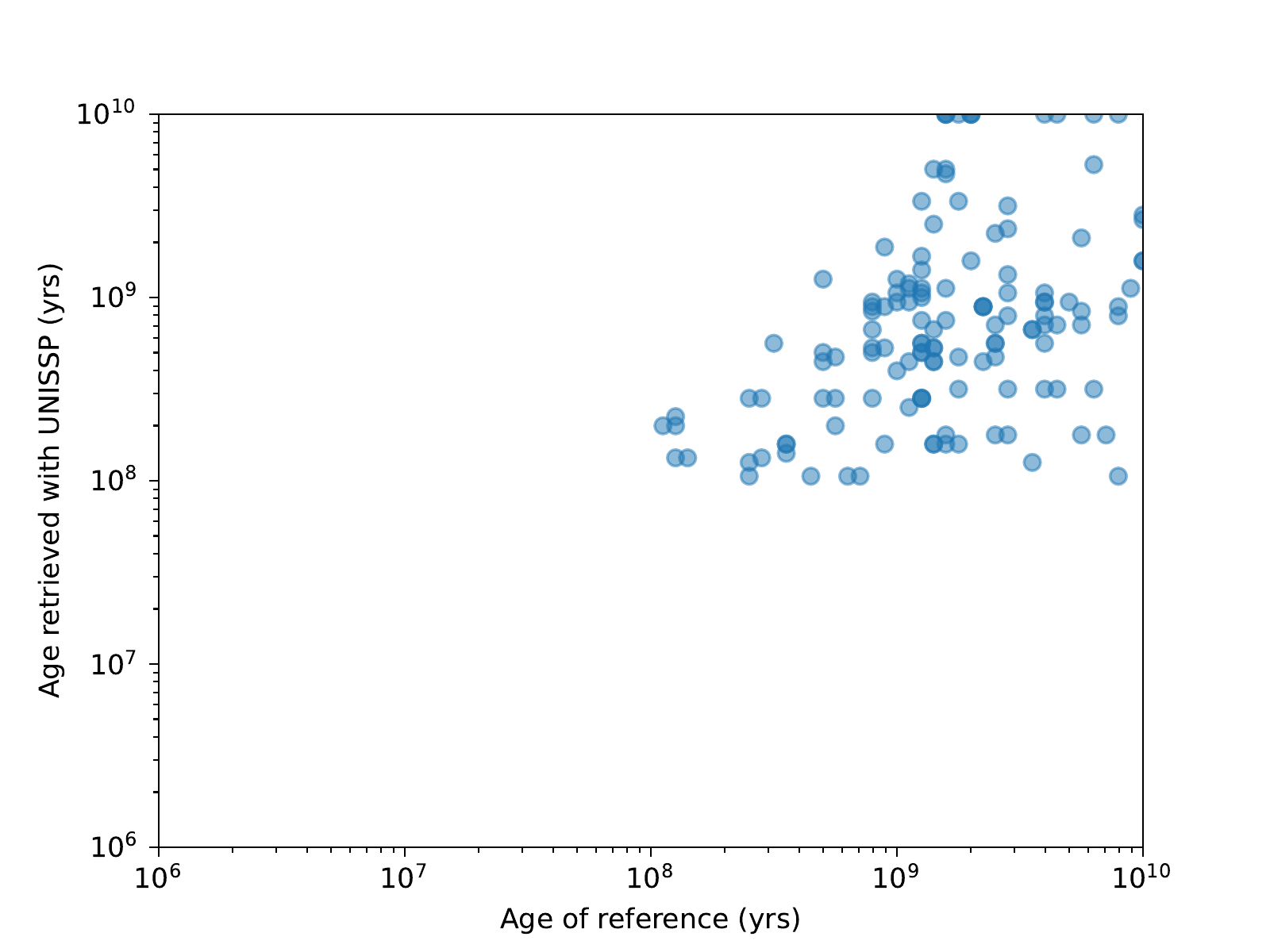}
    \caption{Retrieval of the M33 cluster age \citep{Fan2014} with the UNISSP library based on UBVRI photometry.}
    \label{fig:UNISSP_phot}
\end{figure}

\begin{figure}
    \centering
    \includegraphics[width=0.99\linewidth]{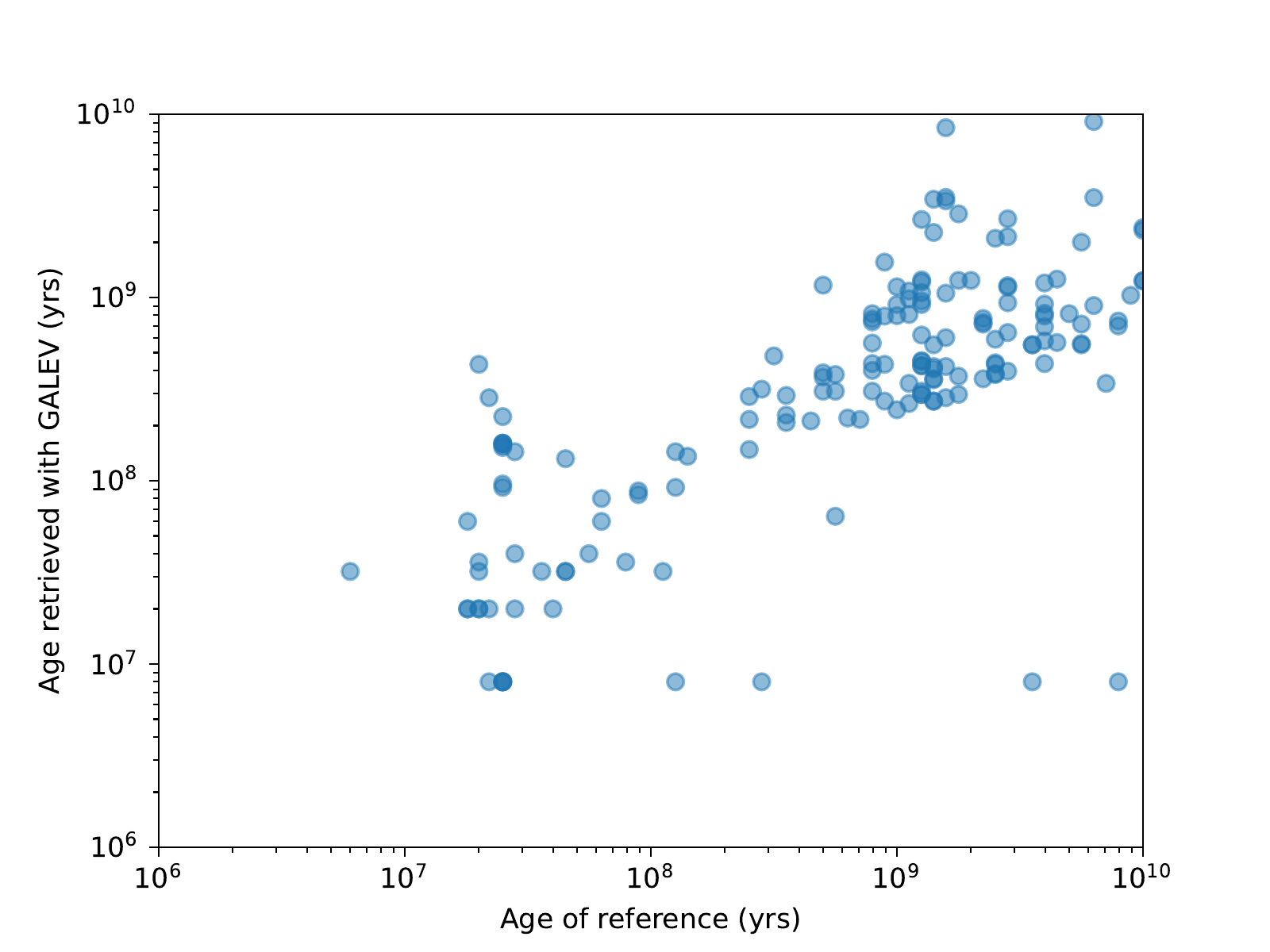}
    \caption{Retrieval of the M33 cluster age \citep{Fan2014} with the GALEV library based on UBVRI photometry.}
    \label{fig:GALEV_phot}
\end{figure}

\begin{figure}
    \centering
    \includegraphics[width=0.99\linewidth]{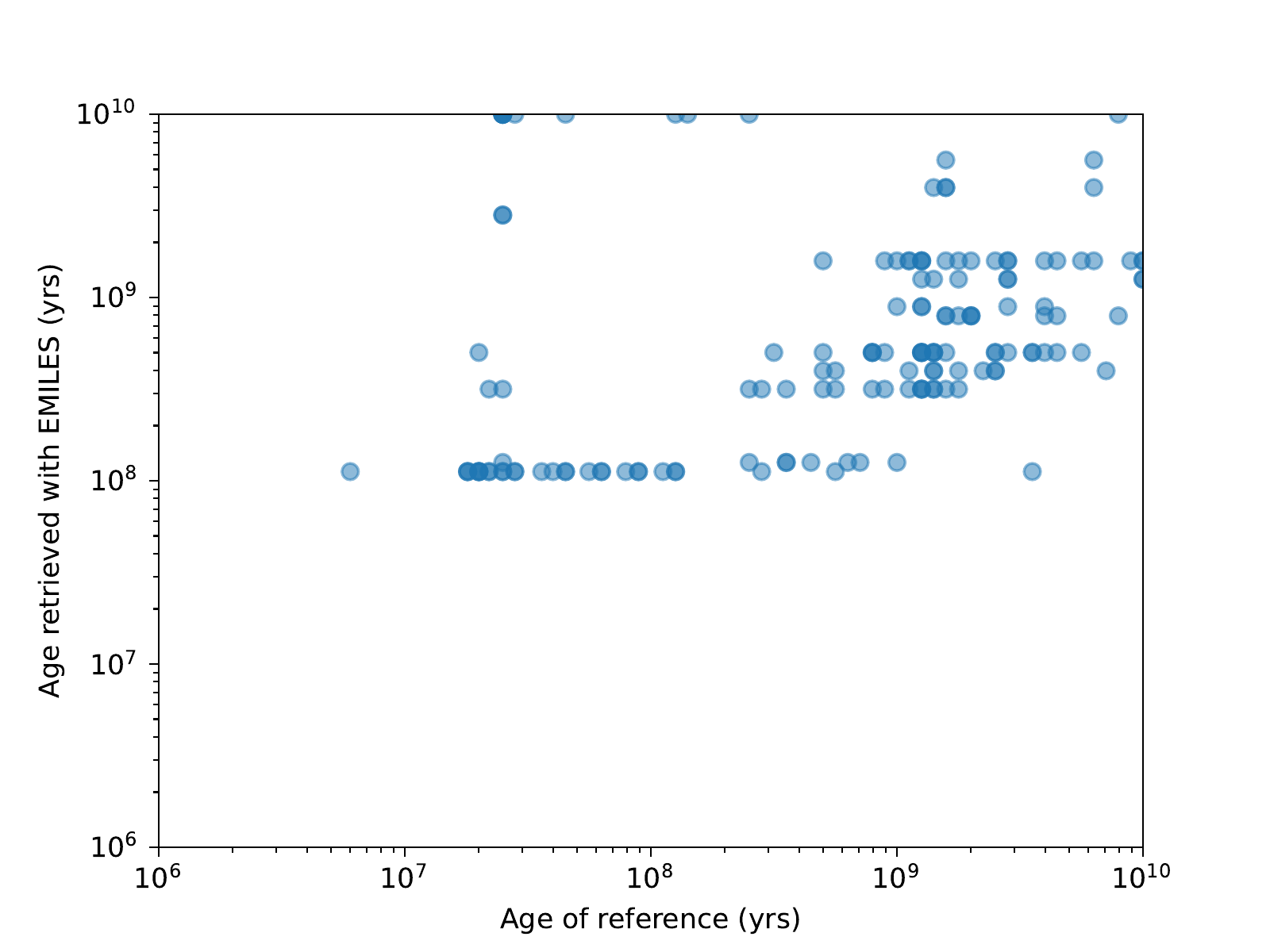}
    \caption{Retrieval of the M33 clusters age \citep{Fan2014} with the EMILES library based on UBVRI photometry.}
    \label{fig:EMILES_phot}
\end{figure}

For each cluster in the catalog, we compared the [(U-B), (U-V), (U-R), (U-I)] colors to the ones predicted by the four libraries, and looked for the age providing the best fit. 
For each library, we present the retrieved age distribution in Figs. \ref{fig:UMISSP_phot}, \ref{fig:UNISSP_phot},
\ref{fig:GALEV_phot}, and \ref{fig:EMILES_phot}, and Table \ref{table_phot} shows the residuals as well as the associated Pearson's correlation coefficient between the published ages and retrieved ones. The results are similar for the four libraries: with only a few exceptions (less than 10 outliers) we are able to distinguish the very young, young, intermediate, and old stellar populations, but with a large spread in the age estimate ($0.6\ dex \lesssim \sigma_{\tau} \lesssim 0.9\ dex$). 

Acknowledging this large dispersion, our two libraries perform similarly to GALEV and EMILES, with a large spread, but are still able to determine the age group. We consider this sanity check to be fulfilled.
\begin{table}

\caption{Residuals and Pearson's correlation coefficient for the age retrieval based on UBVRI photometry with the four spectral libraries.}
\label{table_phot}
\centering
\begin{tabular}{|c||c|c|}
 \hline
 Library & $residuals_{\tau}$ & PCC\\
 \hline
 UMISSP & 0.62 & 0.75 \\
 UNISSP & 0.58 & 0.39 \\
 GALEV & 0.61 & 0.73 \\
 EMILES & 0.62 & 0.38 \\
 \hline
\end{tabular}

\end{table}

\section{Impact of missing stellar templates}
\label{appendix1}
\begin{figure}
    \centering
    \includegraphics[width=0.99\linewidth]{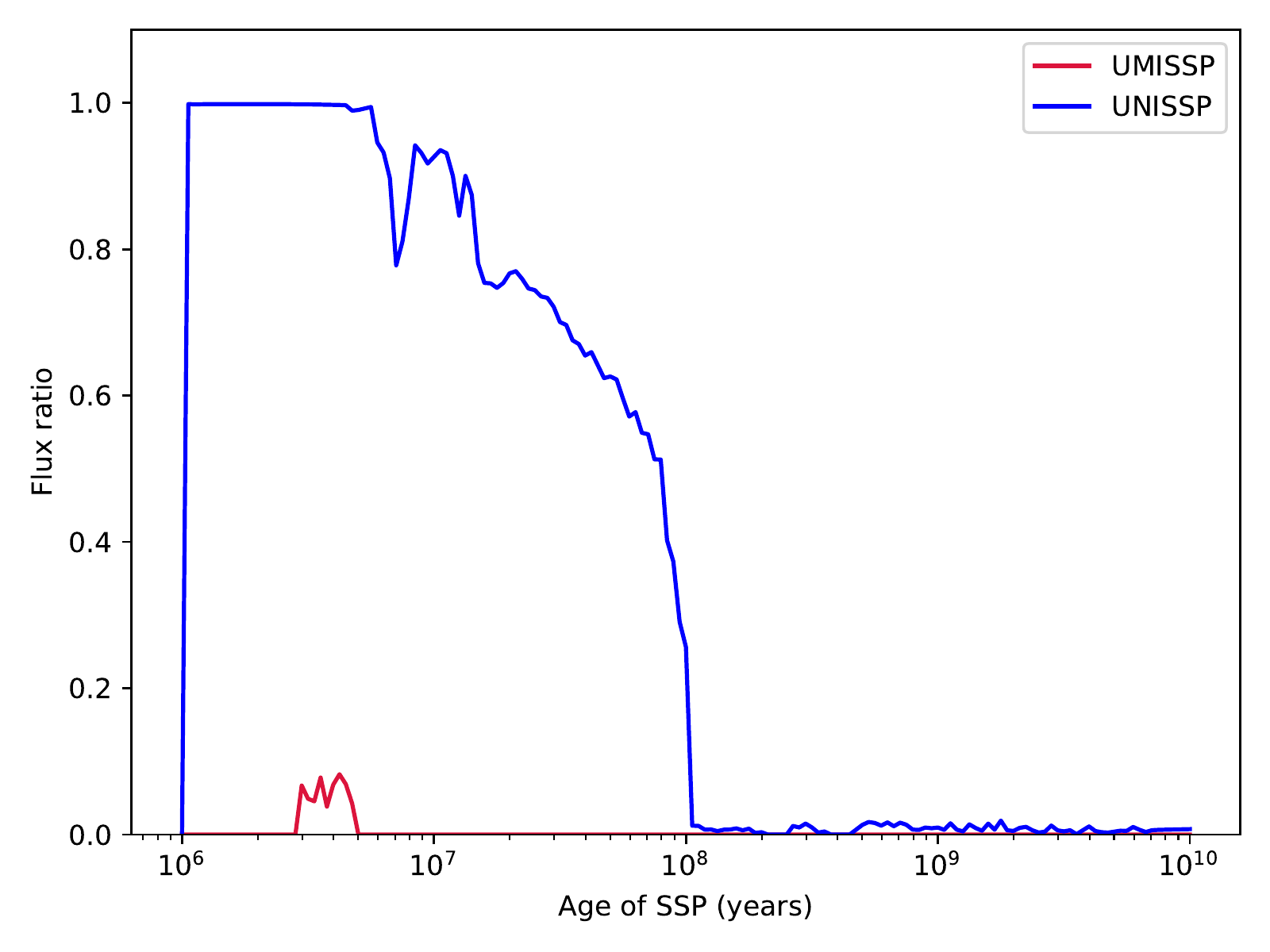}
    \caption{Percentage of the UMISSP and UNISSP flux corresponding to stars with parameters outside the grid ([Fe/H] = 0).}
    \label{mixed_bad}
\end{figure}

In order to estimate the maximum error caused by the missing stellar templates, we measured the bolometric luminosity of the stars whose parameters were outside the model grid, and which were interpolated to the nearest available stellar template. In Fig. \ref{mixed_bad}, we plot the ratio of this flux over the total flux, for UNISSP and UMISSP ([Fe/H] = 0).

With UMISSP, UNISSP, GALEV, and EMILES, we fitted the age of 188 clusters observed in M33 based on their UBVRI photometry. The results are presented in Figs. \ref{fig:UMISSP_phot}, \ref{fig:UNISSP_phot}, \ref{fig:GALEV_phot}, and \ref{fig:EMILES_phot}. In each case, the x-axis corresponds to age of the cluster as published in \citet{Fan2014} and the y-axis corresponds to the fitted age.

\label{appendix2}

\section{Miyamoto Nagai disk}

\label{app:mndisk}
\begin{figure}
    \centering
    \includegraphics[width=0.99\linewidth]{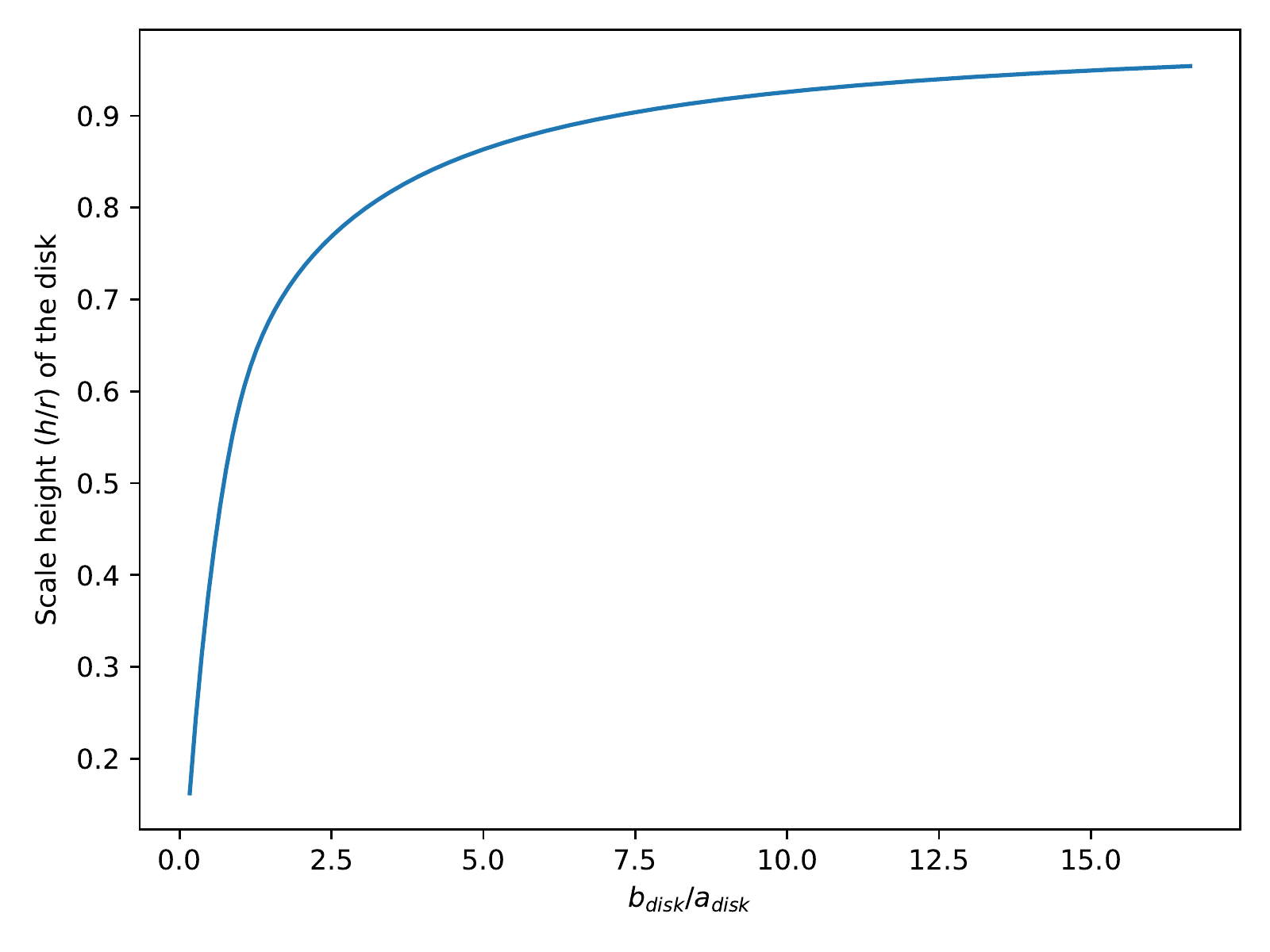}
    \caption{Scale height of the Miyamoto-Nagai disk as a function of $b_{disk}/a_{disk}$ determined by fitting an elongated Gaussian on $\rho_{disk}(a_{disk}, b_{disk})$ density distributions.}
    \label{fig:aoverb}
\end{figure}

In our model, we define the velocity dispersion for each position in the Miyamoto-Nagai disk as purely vertical, and equal in amplitude to the product of the local circular velocity to the scale height of the disk.
However, there is no analytical expression for the scale height of a Miyamoto-Nagai disk because the parameter $b_{disk}$ is not directly providing the size of the disk in the vertical direction. In order to get an estimate of the scale height $(h/r)^*$ of the disk as a function of $a_{disk}$ and $b_{disk}$, we computed a large number of  Miyamoto-Nagai mass distributions with variable $b_{disk}/a_{disk}$, and measured the scale height of the disk by fitting a 2D Gaussian on the mass distribution in a vertical plane passing through the center. The resulting table is presented in Fig. \ref{fig:aoverb}. Then, during our modeling, the $(h/r)^*$ value for any $b_{disk}/a_{disk}$ ratio is interpolated from this table.

\end{appendix}

\end{document}